\documentclass[%
twocolumn,
superscriptaddress,
showpacs,
 amsmath,amssymb,
 aps,
prx,
]{revtex4-1}
 
\usepackage{graphicx}
\usepackage{dcolumn}
\usepackage{bm}
\usepackage{color}
\usepackage{multirow}
\usepackage[normalem]{ulem}
\usepackage{hyperref}

\definecolor{green}{rgb}{0,0.7,0.3}

\begin{document}

\title{
Dirac-type nodal spin liquid revealed by refined quantum many-body solver using neural-network wave function, correlation ratio, and level spectroscopy
}

\author{Yusuke Nomura}
\email{yusuke.nomura@riken.jp}
\affiliation{RIKEN Center for Emergent Matter Science, 2-1 Hirosawa, Wako, 351-0198, Japan}
\author{Masatoshi Imada}
\affiliation{Toyota Physical and Chemical Research Institute, 41-1 Yokomichi, Nagakute, Aichi, 480-1192, Japan}
\affiliation{Research Institute for Science and Engineering, Waseda University, 3-4-1 Okubo, Shinjuku-ku, Tokyo, 169-8555, Japan} 

\date{January 27, 2020}

\begin{abstract}
Pursuing fractionalized particles that do not bear properties of conventional measurable objects, exemplified by bare particles in the vacuum such as electrons and elementary excitations such as magnons, is a challenge in physics. 
Here we show that a machine-learning method for quantum many-body systems that has achieved state-of-the-art accuracy reveals the existence of a quantum spin liquid (QSL) phase in the region $0.49\lesssim J_2/J_1\lesssim 0.54$ convincingly in spin-1/2 frustrated Heisenberg model with the nearest and next-nearest neighbor exchanges, $J_1$ and $J_2$, respectively, on the square lattice. 
This is achieved by combining with the cutting-edge computational schemes known as the correlation ratio and level spectroscopy methods to mitigate the finite-size effects. 
The quantitative one-to-one correspondence between the correlations in the ground state and the excitation spectra found in the present analyses enables the reliable identification and estimation of the QSL and its nature.
The spin excitation spectra containing both singlet and triplet gapless Dirac-like dispersions signal the emergence of gapless fractionalized spin-1/2 Dirac-type spinons in the distinctive QSL phase. 
Unexplored critical behavior with coexisting and dual power-law decays of N\'{e}el antiferromagnetic and dimer correlations is revealed. 
The power-law decay exponents of the two correlations differently vary with $J_2/J_1$ in the QSL phase and thus have different values except for a single point satisfying the symmetry of the two correlations.
The isomorph of excitations with the cuprate $d$-wave superconductors revealed here implies a tight connection between the present QSL and superconductivity.  
This achievement demonstrates that the quantum-state representation using machine learning techniques, which had mostly been limited to benchmarks, is a promising tool for investigating grand challenges in quantum many-body physics.
\end{abstract}


\maketitle

\section{Introduction}

Collective excitations such as magnons and phonons consist of many elementary particles and provide us with fundamental understanding beyond the non-interacting picture, where the spontaneous symmetry breaking and associated Nambu-Goldstone bosons are required in many cases. Fractionalization, on the other hand, offers another route to realize emergent particles manifesting even in the absence of the symmetry breaking and serves as one of the central concepts in modern physics. 
The conventional elementary particles themselves can often be viewed as a bound state of more elementary objects, namely, the fractionalized particles, and such exotic particles emerge through the deconfinement.
A prominent example of the deconfinement occurs in quantum chromodynamics: 
The proton and neutron that had been considered to be elementary particles before have turned out each to be a composite particle of three quarks with fractionalized charges, though quarks are hardly detected in experiments directly because of the confinement. 
In condensed matter, though the electron is an elementary particle in the vacuum, such deconfinement of electrons can be seen at low energies in specific circumstances followed by the ground-state structure of materials.
Consequential emergent fractionalized particles were discovered in examples of polyacetylene soliton~\cite{SuSchriefferHeeger} and fractional quantum Hall states~\cite{PhysRevLett.50.1395}.
The expectation would be that the emergent particles arising from the fractionalization still have particle character as low-energy excitations distinct from the elementary particles in the vacuum and the collective excitations in the symmetry broken states, and then would have novel functions in their many-body states, which may be useful for future applications such as quantum computing.
 
The QSL is a potential platform of such a fractionalization, where suppressed magnetic order by geometrical frustration of the spin interaction is expected to drive the fractionalization. 
The QSL phase was theoretically proposed both through numerical supports and mean-field theories~\cite{Balents_Nature,RevModPhys.89.025003}. 
Experimental efforts also supported the existence~\cite{RevModPhys.89.025003}.  

However, theoretical and experimental efforts have not yet identified and established the nature of fractionalized particles in reality due to their hidden nature and various theoretical difficulties. 
So far, several different types of QSL have been proposed. 
One of the important properties to characterize the QSL is the excitation spectra: 
They are classified, first, by whether the excitations are gapped [as in the cases of gapped $Z_2$ spin liquids (short-ranged resonating valence bond (RVB) states)~\cite{wen91,readsachdev91} and chiral spin liquid~\cite{kalmeyerlaughlin87}], or gapless~\cite{kashima01,morita02,lee05,mizusaki06,nagaosalee06,balents98,senthil00}.
In the gapless case, 
one candidate is the gapless continuum of both of the singlet and triplet in an extended region of the Brillouin zone~\cite{mizusaki06,lee05}, which may arise, for example, 
if spin-1/2 fermionic spinons emerging from the fractionalization constitute a large Fermi surface (or line) as in $U(1)$ spin liquid~\cite{nagaosalee06}. 
Another proposal is the spinon nodal liquid, where a small number of spinon gapless points appear in the Brillouin zone, resulting in the discrete gapless points of the observable spin excitation as well~\cite{balents98,senthil00} (see Fig.~\ref{fig_fractionalization} shown later for illustration). 
At the gapless points, the dispersion may be either linear (Dirac dispersion) or quadratic.

To establish the real existence of the QSL and then narrow down the nature of the QSL, we need to identify excitation spectra connected to experimental indications for a proper Hamiltonian that really accommodates the QSL state. 
However, it remains a challenge because of  highly competing energies of various quantum states.
We need a highly accurate framework for both ground and excited states in a momentum-resolved fashion.

Such high accuracy is offered by a recently developed machine learning method for the ground state. Here, we extend this method to represent both the ground and excited states. 
To be more precise, we employ the restricted Boltzmann machine (RBM) combined with pair-product (PP) states~\cite{PhysRevB.96.205152}.
The RBM+PP method is further supplemented by two independent state-of-the-art numerical procedures, namely the correlation ratio~\cite{PhysRevLett.115.157202} and level spectroscopy~\cite{kiyohide.nomura95} methods, to reach the thermodynamic limit quickly by reducing the finite-size effect. 

We then apply the RBM+PP to a candidate Hamiltonian of the spin-$1/2$ antiferromagnetic (AF) Heisenberg model on the square lattice with the nearest-neighbor and next-nearest-neighbor exchange interactions, $J_1$ and $J_2$, respectively, called the $J_1$-$J_2$ Heisenberg model. We employ two independent analyses to settle down the controversy and obtain firm evidence for the QSL phase:
A finite range of the QSL phase in the region $0.49\lesssim J_2/J_1\lesssim 0.54$ is found.
In the QSL phase, the singlet and triplet excitations are both gapless at four symmetric momenta in support of the nodal Dirac (or quadratic touching) dispersion of the fermionic spinon at $(\pm\pi/2,\pm\pi/2)$ in the Brillouin zone, which brings the coexisting power-law decay of spin-spin and dimer-dimer correlations. 
The isomorphic structure of the gapless excitations of spinons at $(\pm \pi/2,\pm \pi/2)$ with the $d$-wave superconducting state in the cuprate superconductors is suggestive of a mutual profound connection.

\section{\mbox{\boldmath $J_1$}-\mbox{\boldmath $J_2$} Heisenberg model on square lattice}
\label{sec_J1J2model}

The two-dimensional (2D) $J_1$-$J_2$ Heisenberg Hamiltonian reads 
\begin{eqnarray} 
 {\mathcal H} =  J_1 \sum_{ \langle i, j \rangle}  {\bf S}_i  \cdot {\bf S}_j +  J_2 \sum_{ \langle \langle  i, j \rangle \rangle}  {\bf S}_i  \cdot {\bf S}_j, 
\label{J1J2Hamiltonian}
 \end{eqnarray}
where ${\bf S}_i$ is the spin-$1/2$ operator at site $i$, whose $\alpha$ ($\alpha=x$, $y$, $z$) component is  $S_i^\alpha = \frac{1}{2} {\bf c}_i^\dagger \sigma_\alpha {\bf c}_i$ with the electron operator ${\bf c}_i^\dagger  = (c_{i\uparrow}^\dagger,c_{i\downarrow}^\dagger)$ and the Pauli matrix  $ \sigma_\alpha$.
We set $J_1=1$ as the energy unit and we restrict the parameter range as $0 \leq J_2  \leq 1 $. 
$ \langle i, j \rangle$ and $\langle \langle i, j \rangle \rangle$ denote nearest-neighbor and next-nearest-neighbor bonds, respectively.

In this model, the $J_1$ and $J_2$ interactions compete with each other (the former favors the N\'eel-type AF configurations, whereas the latter favors the stripe-type AF configurations).  
Although it is clear that N\'eel- and stripe-type AF phases exist for small and large $J_2$ regions, respectively, around $J_2 = 0.5$, which is the classical boundary between the N\'eel and stripe phases, unconventional quantum phase(s) such as QSL may emerge. 
Despite many theoretical efforts~\cite{Chandra_1988,Capriotti_2001,Zhang_2003,jiang12,Wang_2013,hu13,Qi_2014,gong14,Richter_2015,Morita_2015,Wang_2016,Poilblanc_2017,Haghshenas_2018,Haghshenas_2018,Liu_2018}, 
the intermediate phase(s) of this model is still controversial. 
Indeed, the studies performing a systematic investigation of $J_2$ dependence with modern numerical techniques have proposed different scenarios:
  QSL (either gapless~\cite{hu13,Liu_2018} or gapped~\cite{jiang12}), valence bond solid (VBS) (either columnar~\cite{Haghshenas_2018} or plaquette~\cite{gong14}), or both of them~\cite{Morita_2015,Wang_2018}. 
Deconfined quantum criticality was also proposed instead of the QSL phase~\cite{gong14,Wang_2016}, which is interpreted as the QSL phase shrunk to a point and the fractionalization occurs only at this continuous phase transition point between the two symmetry-broken states. 

To settle the phase diagram after various highly controversial proposals, one needs to satisfy at least the following three requirements:
\begin{enumerate}
\item Systematic investigation on finely-resolved $J_2$ dependence must be performed to establish whether the QSL exists as a phase in a finite $J_2$ interval because the QSL region is not expected to be wide. 
\item Calculation must be highly accurate because quite different states are highly competitive with near degeneracy of the energies.
\item Reliable estimate of the thermodynamic limit to ensure it in the realistic bulk systems. When finite-size systems are studied, methods that are size-independent or have small finite-size effects are required to allow reliable extrapolation to the thermodynamic limit.
\end{enumerate}
Although recent rapid progress in variational numerical methods has contributed to better accuracy, previous studies satisfying all the points hardly exist.
Most of the studies have argued whether the order parameter is finite or not in the thermodynamic limit; however, the order parameter is tiny around the continuous phase transitions, and it is hard to discuss whether the order parameter is really zero or not. 
Also, when finite-size systems are studied, the direct extrapolation of the order parameter has a large finite-size effect.
Exceptionally, Ref.~\cite{Wang_2018} employed the level spectroscopy analysis, which can mitigate the finite-size effect. 
However, it is an indirect method, which speculates phase transitions in the ground state indirectly from the excitation structure. Because there exists no rigorous proof for the one-to-one correspondence of the ground state and excitation structure, one needs to verify in ground-state quantities to settle the highly controversial issue. 

As is detailed below, we employ the RBM+PP method, which offers a unique way of calculating ground state and momentum-space excitation dispersion in a systematically improvable and tractable way, to satisfy the conditions 1 and 2. 
The high accuracy and tractable computational cost of the RBM+PP enable comprehensive correlation ratio (ground-state property) and level spectroscopy (excited-state property) analyses with small finite-size effects.  
To fulfill the condition 3, a crosscheck from the two independent analyses is essential. 

\section{Methods}

\subsection{Machine learning for quantum many-body systems}
Physical properties of many-body systems are governed by the eigenstates of the many-body Hamiltonian.
Therefore, once the eigenstates of the Hamiltonian in Eq.~(\ref{J1J2Hamiltonian}) are known, we can predict the nature of the $J_1$-$J_2$ model precisely. 
However, there is difficulty in obtaining eigenstates because the dimension of the eigenstates grows exponentially as the system size increases. 
In the present case where we consider the $J_1$-$J_2$ Hamiltonian on the $L \times L$ ($=N_{\rm site}$) lattice with the periodic boundary condition, we cannot obtain the exact wave function when $N_{\rm site} \gtrsim 50$.
However, by using machine learning techniques, we can compress the data of eigenstates and approximate the wave functions accurately with a finite number of parameters. 

Here, we employ a newly developed machine learning method, RBM+PP~\cite{PhysRevB.96.205152}, to obtain accurate representations for both the ground and excited states. 
The RBM is a type of artificial neural network having two (visible and hidden) layers~\cite{RBM_Smolensky}. 
Using the machine learning technique, one can construct accurate many-body wave functions, which are systematically improvable toward the exact solution~\cite{Carleo602}. 
Indeed, it has been shown both theoretically and numerically that the RBM variational state flexibly describes a variety of quantum states~\cite{Carleo602,PhysRevX.7.021021,PhysRevB.96.195145,PhysRevB.97.085104,PhysRevX.8.011006,1751-8121-51-13-135301,PhysRevB.97.195136,PhysRevB.99.155136,Huang_arXiv,Vieijra_2020,Nomura_2020,Carleo_2019},
including the states exhibiting the volume-law entanglement entropy~\cite{PhysRevX.7.021021,PhysRevB.97.085104}, which is advantageous to represent not only the ground state but also the excited states. 
Indeed, the RBM is shown to accurately describe excited states of quantum spin Hamiltonians~\cite{Choo_2018,Nomura_2020}, for which existing numerical methods often encounter numerical difficulties. 
Meanwhile, the PP state (called ``geminal" in quantum chemistry) is represented by fermion wave functions, which can also accommodate volume-law entanglement. 
The PP state mapped onto bosonic spin space can represent RVB states~\cite{ANDERSON1196}, serving as a powerful starting point of the ground state approximation for the quantum spin systems~\cite{PhysRevB.42.6555}. 
The combined wave function, RBM+PP, inherits advantages of both and acquires much better accuracy than those achieved by either of the RBM or PP state separately~\cite{PhysRevB.96.205152}. 
By the RBM+PP method with quantum number projections (see below), we can calculate momentum resolved excitations.

The RBM+PP wave function $\Psi(\sigma) = \langle \sigma | \Psi \rangle$ with $|\sigma\rangle=\prod_{i}c_{i\sigma}^{\dagger}|0\rangle$ is given by (we neglect normalization factor)~\cite{PhysRevB.96.205152} 
\begin{eqnarray}
\Psi ( \sigma ) =  \phi_{\rm RBM} (\sigma) \psi_{\rm PP} (\sigma)
\label{eq:Psi_sigma}
\end{eqnarray}
for each spin configuration $\sigma = (\sigma_1, \sigma_2, \ldots, \sigma_{N_{\rm site}} )$ with $\sigma_i = 2 S_i^z = \pm 1$.
The number of sites is given by $N_{\rm site} =L \times L$ and the periodic boundary condition is assumed. 
The RBM part is given by (we omit irrelevant bias term  on the physical spins)
\begin{eqnarray}
\phi_{\rm RBM} (\sigma)   =    \sum_{ \{ h_k\}}  \exp \biggl ( \sum_{i,k} W_{ik}  \sigma_i h_k +\sum_{k} b_k h_k \biggr ) 
 \label{eq:RBM(sigma)}
\end{eqnarray}
with the spin state of hidden units $h_k = \pm1$, the interaction between physical and hidden variables $W_{ik}$, and the bias on the hidden variables $b_k$.   
The number of hidden units is taken to be $16$. 
The sum over hidden variables can be evaluated analytically and Eq. (\ref{eq:RBM(sigma)}) can be efficiently computed as $\phi_{\rm RBM} (\sigma) =  \prod_k 2 \cosh \bigl ( b_k + \sum_i W_{ik} \sigma_i  \bigr )$.
To make it possible to express the sign change of the wave function, we take the $b_k$ and $W_{ik}$ variational parameters to be complex. 
The PP state mapped onto spin systems reads
\begin{eqnarray}
\bigl | \psi_{\rm PP } \bigl \rangle = P_{\rm G} \biggl( \sum_{i,j}  f_{ij}^{\uparrow \downarrow} c^\dagger_{i \uparrow}  c^\dagger_{j\downarrow}   \biggr ) ^ {  N_{\rm site}/2 } \bigl |  0 \bigr \rangle
\end{eqnarray}
with real variational parameters $f_{ij}^{\uparrow \downarrow}$.
$\psi_{\rm PP}(\sigma)$ in Eq.~(\ref{eq:Psi_sigma}) is related as $\psi_{\rm PP}(\sigma)\equiv\langle \sigma | \psi_{\rm PP}\rangle$. 
Here, $P_{\rm G} = \prod_i ( 1 - n_{i\uparrow} n_{i\downarrow})$ with $n_{i \uparrow} = c^\dagger_{i \uparrow} c_{i \uparrow}$ and $n_{i \downarrow} = c^\dagger_{i \downarrow} c_{i \downarrow}$ is the Gutzwiller projection prohibiting double occupancy. 

We optimize the variational parameters $\{ b_k, W_{ik},  f_{ij}^{\uparrow \downarrow} \}$ to minimize the energy 
$ E =  \frac{ \langle \Psi | { \mathcal H }  | \Psi  \rangle  } { \langle \Psi | \Psi \rangle  }$. 
The energy is a highly nonlinear function with respect to the parameters  $\{ b_k, W_{ik},  f_{ij}^{\uparrow \downarrow} \}$. 
Therefore, by interpreting the energy as a loss function, the task of obtaining the lowest-energy state can be recast as a machine-learning task, namely
a high-dimensional optimization problem of the highly nonlinear function (RBM+PP) using the highly nonlinear loss function (energy)~\cite{Melko_2019}. 
The details of the optimization method and the calculation conditions can be found in Appendix~\ref{Appendix_method_detail}.

\subsection{Strategy to overcome numerical challenges}
Various competing controversial scenarios have been proposed for the phase diagram and the nature of possible QSL as we mentioned above. The machine learning only is, despite its crucial importance, not enough to resolve these controversies. 
In fact, even when we obtain accurate representations of quantum states by the machine learning, (i) another challenge is how to reach quick convergence to the thermodynamic limit from available finite-size results (condition 3 listed in Sec.~\ref{sec_J1J2model}). 
Furthermore, provided that the QSL phase exists, the next challenge is to elucidate its nature; (ii) it is essential to estimate the excitation gap structure and momentum resolved dispersion accurately. 
To overcome the challenge (i), the present paper employs an unprecedented combination of two methods and one supplementary analysis together and
reaches quantitative agreements, which ensures the accuracy because the two methods are originally independent of each other. 
As a computational method to identify the quantum phases, this is the first attempt to use such combinations, and it successfully establishes a way to obtain the accurate phase diagram, which may serve as the standard method in the future. 
The first method is the correlation ratio method~\cite{PhysRevLett.115.157202}, which utilizes the ground state properties (see Sec.~\ref{sec_method_correlation_ratio}).
The second is the level spectroscopy~\cite{kiyohide.nomura95},
which detects the signature of the phase transition in the excitation spectra (see Sec.~\ref{sec_method_level_spectroscopy}).
Both methods show small finite-size effects and quickly converge to the thermodynamic limit. 
These methods were developed independently and, in fact, measure the excited and ground-state properties, respectively, which are originally independent. However, the important point is that they have the one-to-one correspondence, conceptually similar to the fluctuation-dissipation theorem and Kubo formula, between the equilibrium and non-equilibrium excited states. 
Such correspondence and match in the calculated results help to ensure the reliability of the phase diagram. 
Further, the obtained phase boundary is supported by the standard finite-size scaling method thanks to the universal scaling relations (see Sec.~\ref{sec_method_finite_size}).
For (ii), we use quantum number projection to reach the accuracy on the spectroscopy level~\cite{Mizusaki} (see Sec.~\ref{sec_qp_proj}). 
Here, we address the advantages of employing these methods.

\subsubsection{Correlation ratio} \label{sec_method_correlation_ratio}
Correlation ratio $R$ quantifies how sharp the structure factor peak is.
$R$ is given by $R=1-S({\bf Q}+\delta {\bf q})/S({\bf Q})$~\cite{PhysRevLett.115.157202,PhysRevLett.117.086404}, 
where $S({\bf q})$ is the structure factor, ${\bf Q}$ is the peak momentum, and ${\bf Q}+\delta {\bf q}$ is the neighboring momentum. 
In the case of the square lattice,  $\delta {\bf q}= (2\pi / L ,0)$ and $(0,2\pi / L )$.
We see that the $R$ value approaches 1 (0) when the peak becomes sharp  (broad). 
Therefore, with increasing system size, $R$ scales to 1 in the ordered phase with delta-function Bragg peak and 0 in the disordered phase.
The crossing point of $R$ curves for different system sizes does not depend sensitively on system size. Thus it is suitable for an accurate estimate of the phase boundary between ordered and disordered phases in the thermodynamic limit~\cite{PhysRevLett.115.157202,PhysRevLett.117.086404}. 

We examine $R$ for both spin-spin and dimer-dimer correlations to detect N\'eel-AF and VBS transition points, respectively. 
The spin-spin correlation is given by $C_{\rm s} ( {\bf r}_i - {\bf r}_j) = \langle {\bf S}_i  \cdot {\bf S}_j \rangle $.
The dimer-dimer correlation is defined as $C_{{ \rm d}_\alpha} ( {\bf r}_i - {\bf r}_j) = \langle D^\alpha_i  D^\alpha_j  \rangle  - \langle D^\alpha_i  \rangle \langle D^\alpha_j  \rangle$ with the dimer operator $ D^\alpha_i = {\bf S}_i  \cdot {\bf S}_{i + \hat{\alpha} }$ on the nearest-neighbor bonds for the $\alpha$-direction ($\alpha=x$, $y$).
Hereafter, the subscripts ``s",  ``d$_x$", and ``d$_y$" are used for spin-spin, dimer-dimer ($\alpha =x$ and $\alpha =y$) correlations, respectively. 
Then, the structure factor is calculated from $S_{{ \rm \gamma}} ( {\bf q} ) =   \frac{1} {N_{\rm site} }  \sum_{i,j} C_{{ \rm \gamma}} ( {\bf r}_i - {\bf r}_j) e ^{i {\bf q} \cdot( {\bf r}_i - {\bf r}_j )}$  with $\gamma = $ s,  d$_x$, and d$_y$.
The two correlation ratios $R_{\rm N\acute{e}el}$ and $R_{\rm VBS}$ are defined from  $S_{{ \rm s}} ({\bf q})$ and  $S_{{\rm d}_x} ({\bf q})$ [or equivalently $S_{{\rm d}_y} ({\bf q})$] to determine the N\'eel-AF and VBS transition points, respectively. 
Close to the N\'eel-AF phase, the peak momentum is ${\bf Q} = (\pi,\pi)$ for $S_{\rm s} ({\bf q})$.  
For VBS, ${\bf Q} = (\pi,0)$ for $S_{{\rm d}_x} ({\bf q})$ and ${\bf Q} = (0,\pi)$ for $S_{{\rm d}_y} ({\bf q})$.

\subsubsection{Level spectroscopy} \label{sec_method_level_spectroscopy}
Quantum phases are characterized by their unique structure in excitation spectra. 
At finite sizes, if the phases are different, low-lying excitations will be characterized by different quantum numbers. 
Therefore, the transition point can be estimated by the size extrapolation of the crossing point of the low-lying excitation energies~\cite{kiyohide.nomura95}. 
This level spectroscopy method is known to have small system size dependence as well. 
Indeed, it has played an important role in precisely determining the Berezinskii-Kosterlitz-Thouless transition point for the sine-Gordon model~\cite{kiyohide.nomura95}. This method offers an analysis completely different but complementary to the correlation ratio method.

\subsubsection{Quantum number projection}
\label{sec_qp_proj}

The eigenstates of the Hamiltonian in finite size systems are labeled by quantum numbers.
By optimizing the RBM+PP wave function for each quantum number sector, we can obtain both the ground state and low-lying excited states. 
We apply total-momentum and spin-parity projections to the RBM+PP wave functions to specify the quantum number~\cite{Mizusaki}:
\begin{eqnarray}
 \Psi_{\bf K}^{S_\pm} ( \sigma ) = \sum_{{\bf R} }  e^{ - i {\bf K}  \cdot {\bf R} }   [   \Psi ( T_{\bf R}   \sigma )  \pm  \Psi (- T_{\bf R}   \sigma )   ] 
 \label{Eq.quantum_proj}
 \end{eqnarray}
(double sign in the same order). 
 $S_+$ ($S_-$) indicates even (odd) spin parity corresponding to even (odd) values of the total spin $S$. 
${\bf K}$ is the total momentum. $T_{\bf R}$ is a translation operator shifting all the spins by ${\bf R}$. 
For each quantum number sector, we optimize the RBM+PP wave function to obtain the lowest-energy state. 
Although the spin-parity projection can only distinguish whether $S$ is even or odd, 
we always obtain a singlet state for the even $S$ sector and a triplet state for the odd $S$ sector (we confirm it by calculating $S$ expectation value for the obtained states). 
This is because the singlet (triplet) state is the lowest-energy state for each even (odd) $S$ sector.  
We note that the quantum number projection is helpful not only to distinguish quantum numbers but also to lower the variational energy~\cite{doi:10.1143/JPSJ.77.114701}.

The ground state is given for ${\bf K} = (0,0)$ and even $S$ sector. 
The energies for other quantum number sectors measured from the ground state energy determine the excitation spectra. 
Then, we can obtain singlet and triplet excitations separately with momentum resolution. 
Exceptionally, we need special treatments to obtain $S=0$ excited state at ${\bf K} = (0,0)$ and $S=2$ excited states, which are described in detail in Appendix~\ref{Appendix_excited}. 
As we mentioned above, the flexible representability of the RBM+PP gives accurate representations not only for ground states but also for excited states. 
The accurate estimate of momentum-resolved excitation gaps enables us to perform the above-mentioned level spectroscopy and also to elucidate the nature of the QSL phase.

\subsection{Finite-size scaling method}
\label{sec_method_finite_size}
Near the quantum critical point, the susceptibility $\chi$ at the ordering wave vector ${\bf Q}$ in finite-sized systems follow the following finite-size scaling form~\cite{Fisher_Barber} 
\begin{equation}
\frac{\chi(t,{\bf Q},L)}{L^{\gamma/\nu}}=f_{\chi}(L^{1/\nu}t),
\end{equation} 
where the universal scaling function $f_{\chi}$ appears with the correlation length exponent $\nu$ and the susceptibility exponent $\gamma$. 
Here, $t$ assumed to satisfy $t\ll 1$ is the dimensionless distance to the critical point.
In the present case $t=(J_2-J_2^{\rm N\acute{e}el}) /J_1$ or $t=(J_2-J_2^{\rm VBS} )/J_1$.  Through the relation between $\chi$ and the structure factor $S(t,{\bf Q},L)$ given by $\chi(t,{\bf Q},L) \sim S(t,{\bf Q},L)L^z$ with the dynamical exponent $z$, we find that
the squared order parameter $m^2=S(t,{\bf Q},L)/L^d$ for the $d$-dimensional system follows
\begin{equation}
m^2 L^{d+z-2+\eta} =f_{\chi}(L^{1/\nu}t),
\end{equation} 
if the Fisher's scaling relation $\gamma/\nu=2-\eta$ holds for $\eta$ associated with the anomalous dimension characterized by the power-law decay of the correlation, $C({\bf r})\sim 1/r^{d+z-2+\eta}$ for distance $r = |{\bf r}|$ at the critical point.
Then the finite-size scaling plot should exhibit the universal scaling function $f_{\chi}$.

\section{Results}

First, we check the accuracy of the RBM+PP method in analyzing the $J_1$-$J_2$ Heisenberg model (see Appendix~\ref{Appendix_Benchmark}).
We have confirmed that the RBM+PP achieves state-of-the-art accuracy not only among machine-learning-based methods but also among all available numerical methods. 
Indeed, the RBM+PP wave function marks the best precision for the ground state calculations among the compared methods for the $8 \times 8$ and $10\times 10$ lattices (see Fig.~\ref{fig_ene_comparison} and Table~\ref{table:ene_comparion_10x10}). 
We have also found that the RBM+PP represents excited states with unprecedented accuracy (Fig.~\ref{fig_ex_ene_6x6}).

Also, in our RBM+PP method, the computationally most demanding part is coming from the PP part, and the neural-network (RBM) part offers an efficient way of improving accuracy without increasing the scaling of computational cost, i.e., as compared to the PP only calculations with the computational cost of ${\mathcal O}(N_{\rm site}^3)$, the computational cost increases only by ${\mathcal O}(1)$. 
This is in contrast to the Lanczos method, which is also used to improve the variational energy (see Appendix~\ref{Appendix_Benchmark}): if we apply the $p$-th order Lanczos step to improve the PP only calculations, it increases the computational cost by ${\mathcal O}(N_{\rm site}^p)$, and hence the total computational cost of the Lanczos-applied PP calculations scales as ${\mathcal O}(N_{\rm site}^{p+3})$.  
Although we calculated the ground state and various excited states independently, a tractable computational-cost scaling of the RBM+PP method allowed us to perform numerous independent calculations for large system sizes within given computational resources. 
Thus obtained high-quality data contribute to a reliable determination of the phase diagram consistently from both ground-state and excitation analyses (see below). 

\begin{figure}[tb]
\begin{center}
\includegraphics[width=0.48\textwidth]{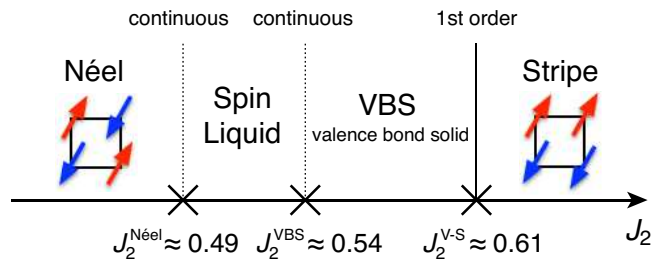}
\caption{
Ground-state phase diagram of square-lattice $J_1$-$J_2$ Heisenberg model ($J_1=1$) obtained by the RBM+PP method.}
\label{fig_phase_diagram}
\end{center}
\end{figure}

\subsection{Ground-state phase diagram}
The RBM+PP method combined with the state-of-the-art numerical techniques convincingly uncovers the phase diagram of the $J_1$-$J_2$ Heisenberg model as shown in Fig.~\ref{fig_phase_diagram}.
In the small (large) $J_2$ region, the N\'eel-type (stripe) AF long-range order appears as in the classical phase diagram. 
In between these two phases, nonmagnetic ground states, QSL and VBS, are found in the region $J_2^{\rm N\acute{e}el}\approx 0.49\le J_2 \le J_2^{\rm VBS} \approx 0.54$ and $J_2^{\rm VBS}\approx 0.54\le J_2 \le J_2^{\rm V\mathchar`-S} \approx 0.61$, respectively.  
Whereas VBS breaks lattice symmetry, QSL does not break any. 
Clearly and notably, QSL is stabilized in a finite region of $J_2$ around $J_2 = 0.5$.  
The phase transition between VBS and stripe-AF at $J_2^{\rm V\mathchar`-S}$ is of 1st order, which is characterized by the kink in the ground state energy, while the other two transitions are continuous (Fig.~\ref{fig_GS_ene} in Appendix~\ref{Appendix_data}). 
Below, we describe the procedure to determine the continuous phase transition points.

\subsubsection{Phase boundary determined by correlation ratio} \label{sec_results_correlation_ratio}
\begin{figure}[tb]
\begin{center}
\includegraphics[width=0.48\textwidth]{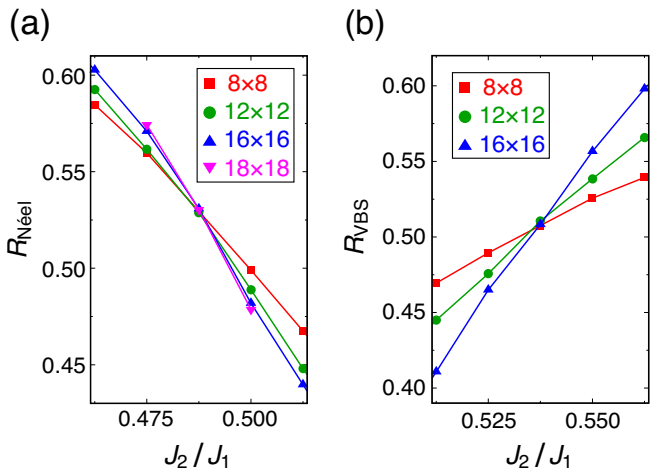}
\caption{
System-size dependence of correlation ratio for (a) spin-spin and (b) dimer-dimer correlations, which are used to detect the phase boundary of N\'eel-AF and VBS, respectively.
In (a), the $18\times18$ data are added to reinforce the result.}
\label{fig_ratio_crossing}
\end{center}
\end{figure}

Results for the correlation ratios, $R_{\rm N\acute{e}el}$ and $R_{\rm VBS}$, are shown in Figs.~\ref{fig_ratio_crossing}(a) and \ref{fig_ratio_crossing}(b), respectively 
(see Figs.~\ref{fig_AF_structure_factor} and \ref{fig_VBS_structure_factor} in Appendix~\ref{Appendix_data} for the raw data of correlation functions). 
We see clear crossings of curves for three sizes at nearly the same points at $J_2 = J_2^{\rm N\acute{e}el} \approx 0.49$ for $R_{\rm N\acute{e}el}$ and at $J_2 = J_2^{\rm VBS} \approx 0.54$ for $R_{\rm VBS}$. 
This standard procedure strongly supports that the two transitions associated with the N\'eel-AF and VBS ordering take place at the different points close to these system-size independent crossings.  
It supports the existence of an intermediate phase without any long-range ordering, i.e., QSL phase in the range $0.49 \lesssim J_2 \lesssim 0.54$ 
(see Appendix~\ref{Appendix_ratio_size_dep} for the discussion of the system-size dependence of the crossing points).

\begin{figure*}[tb]
\begin{center}
\includegraphics[width=0.95\textwidth]{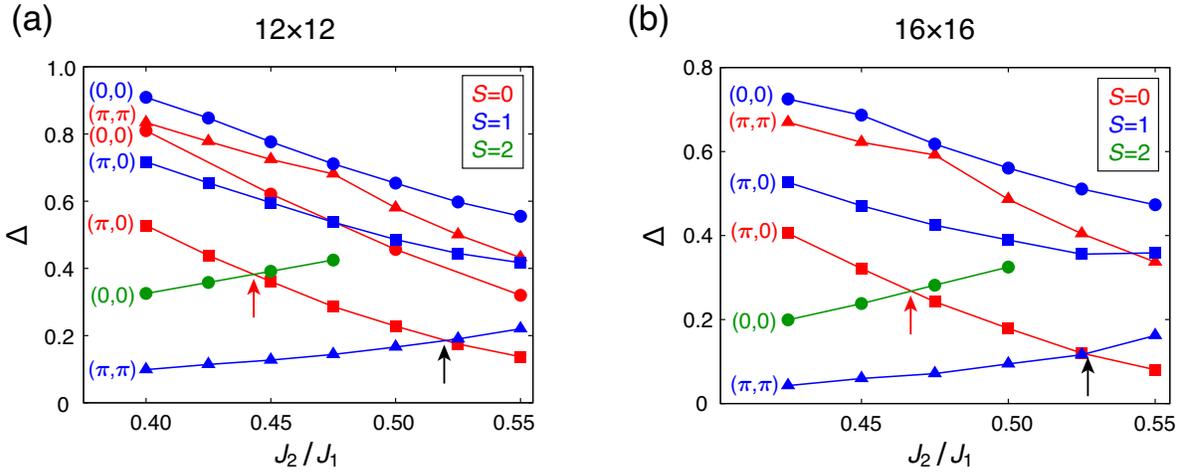}
\caption{
Low-lying excitation energies for $J_1$-$J_2$ Heisenberg model for (a) $12\times 12$ and (b) $16\times 16$ lattices.
The red and black arrows indicate singlet-quintuplet and singlet-triplet level crossings, respectively.}
\label{fig_level_cross_1}
\end{center}
\end{figure*}

\begin{figure}[tb]
\begin{center}
\includegraphics[width=0.48\textwidth]{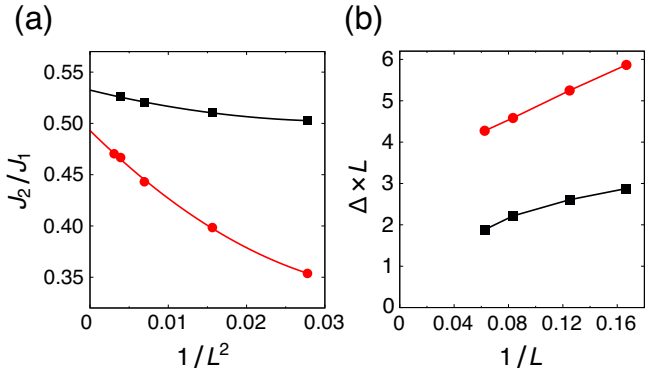}
\caption{
(a) System-size dependence of singlet-quintuplet (red dots) and singlet-triplet (black squares) level crossings indicated by red and black arrows in Fig.~\ref{fig_level_cross_1}. 
The extrapolation to the thermodynamic limit is done by the polynomial fit 
$a + b / L^2 + c/L^4$ (solid curves). 
(b) System-size dependence of the excitation gap $\Delta$ at the two level crossings. 
For the singlet-quintuplet level crossing in (a), the $18\times18$ data are added to corroborate the result.
}
\label{fig_level_cross_2}
\end{center}
\end{figure}

\subsubsection{Phase boundary determined by level spectroscopy} \label{sec_results_level_spectroscopy}
The level spectroscopy method was applied to the 2D $J_1$-$J_2$ Heisenberg model before~\cite{Wang_2018}.  
They interpreted the crossing between the lowest singlet and triplet excitations as the VBS-order boundary, following Ref. \onlinecite{PhysRevB.94.144416}.
In addition, they found the singlet-quintuplet crossing and interpreted it as a signal of the disappearance of the AF long-range order, because the transition from the AF long-range order to quasi-long-range order in one-dimensional Heisenberg model with long-range interaction shows a similar behavior~\cite{Wang_2018,PhysRevLett.104.137204}.
These two crossings extrapolated to $L \rightarrow \infty$ limit gave different $J_2$ values: 
$J_2= 0.463(2)$ and $J_2 = 0.519(2)$ for the singlet-quintuplet and singlet-triplet crossings, respectively.

To critically crosscheck the consistency with the above correlation ratio result, we also reexamine the level spectroscopy analysis as a complementary check. We here enjoy the advantage of the momentum resolution in addition (contrary to Ref.~\onlinecite{Wang_2018}). 
Figure~\ref{fig_level_cross_1} shows $J_2$ dependence of the excitation energies $\Delta$ for sizes (a) $12\times12$ and (b) $16\times16$ at high-symmetry momenta. 
The singlet-quintuplet and singlet-triplet crossings signaling the AF-QSL and QSL-VBS transitions, respectively, are highlighted by arrows.
The size extrapolation of the crossing points is shown in Fig.~\ref{fig_level_cross_2}(a). 
We use $L^{-2}$ scaling as in Refs.~\onlinecite{Wang_2018} and \onlinecite{PhysRevB.94.144416}.
The extrapolated thermodynamic values are $J_2= 0.493(2)$ and $J_2 = 0.532(2)$ for the singlet-quintuplet and singlet-triplet crossings, respectively. 
The values are close to those of Ref.~\onlinecite{Wang_2018} above. Quantitative differences may well be ascribed to the smaller system sizes calculated in Ref.~\onlinecite{Wang_2018} than ours.
As for the singlet-triplet crossing, our result is also consistent with a more recent estimate by the variational Monte Carlo (VMC) method, which gives $J_2= 0.542(2)$~\cite{Ferrari_2020}.

More importantly, our phase boundary estimated by the level spectroscopy has a striking quantitative agreement with the correlation ratio result described above. 
It is of great significance to see the one-to-one correspondence between the ground-state phases and the excitation structures.
We then safely conclude that a finite QSL region around $J_2 = 0.5$ emerges (see Supplementary Note 1 in Appendix~\ref{Appendix_Notes} for additional noteworthy features found in the level spectroscopy). 

Figure~\ref{fig_level_cross_2}(b) further shows the size dependence of the excitation gap $\Delta$ at the crossing points. 
$\Delta \times L$ seems to converge at a finite value as $L \rightarrow \infty$ for both crossings. 
Therefore, the two critical points corresponding to AF-QSL and QSL-VBS transitions become gapless in the thermodynamic limit with the scaling $\Delta \propto 1/L$.

\begin{figure}[tb]
\begin{center}
\includegraphics[width=0.49\textwidth]{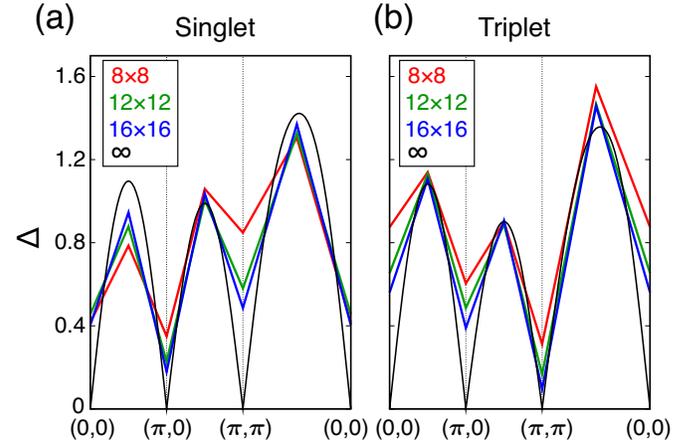}
\caption{
Low-lying excitation in the QSL phase.
(a) Singlet and (b) triplet excitation gap along the symmetric line in the Brillouin zone at $J_2 = 0.5$. 
On top of the high-symmetry ${\bf K}$ points $(0,0)$, $(\pi,0)$, and $(\pi,\pi)$, the excitations at intermediate points $(\pi/2,0)$, $(\pi,\pi/2)$, and $(\pi/2,\pi/2)$ are calculated. 
Black curves are expected dispersions in the thermodynamic limit (see text). 
}
\label{fig_excitation_spectrum}
\end{center}
\end{figure}

\begin{figure*}[tb]
\begin{center}
\includegraphics[width=0.95\textwidth]{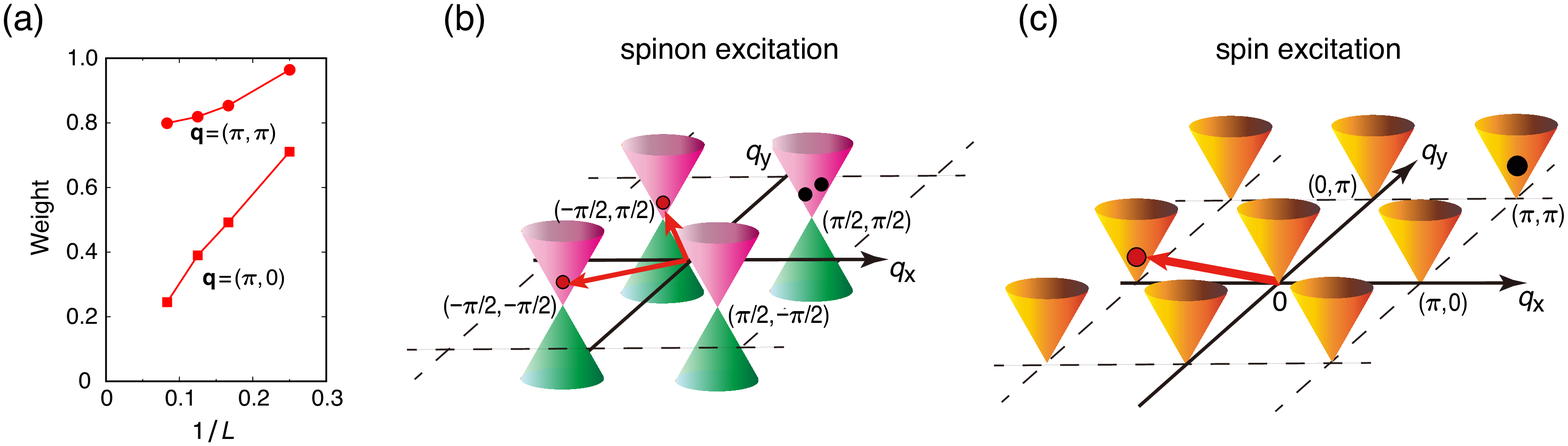}
\caption{
(a) Weight of lowest branch in the dynamic spin structure factor for ${\bf q} = (\pi,0)$ and $(\pi,\pi)$ for $J_2/J_1=0.5$. 
At each ${\bf q}$ point, the weight is normalized by the total spectral weight $\int_0^\infty d \omega S_{\rm s}({\bf q}, \omega)$.  
(b) Schematic picture for plausible spinon dispersion around gapless points $(\pm\pi/2,\pm\pi/2)$, illustrated both for particle (pink) and hole (green) sides above and below the spinon Fermi energy. 
Two examples of two spinon excitations (two red and two black circles) are illustrated (see below). 
(c) The observable spin excitation is constructed from the two spinon excitations, which generates the gapless points at $(\pi,0), (0,\pi), (0,0)$ and $(\pi,\pi)$. For instance, the red circle with the momentum around $(-\pi,0)$ is constructed from the two spinon excitations shown as the small red circles with the momenta around $(-\pi/2,\pi/2)$ and $(-\pi/2,-\pi/2)$ in (b). 
The black circle is another example of spin excitation originated from the two spinon excitations shown as the small black circles in (b). 
Continuum incoherent spin excitations inside the cones are generated from the combinations of the two spinon excitations on the pink or green cone surfaces in (b).
}
\label{fig_fractionalization}
\end{center}
\end{figure*}

\subsection{Excitation spectrum in quantum spin liquid phase}
\label{sec_results_gapless}
As we see in Fig.~\ref{fig_level_cross_2}(b), the singlet excitation with ${\bf K} = (\pi,0)$ and $(0,\pi)$ becomes gapless at both AF-QSL and QSL-VBS critical points, 
implying that it is gapless through the QSL region sandwiched by these two critical points.  
In the QSL phase, the triplet excitation at ${\bf K} = (\pi,\pi)$ is the lowest excited state in finite-size systems [lower than the gapless singlet at $(\pi,0)$] lending support to the vanishing gap also for $(\pi,\pi)$ triplet in the thermodynamic limit.
By the excitation involving the triplet at $(\pi,\pi)$ and the singlet at $(\pi,0)$, one can construct the triplet $(0,\pi)$, which must be gapless if these two elementary excitations are excited far apart in the thermodynamic limit, even when they are repulsively interacting. 
In a similar way, one can construct a gapless singlet excitation at $(\pi,\pi)$ and $(0,0)$. 
Therefore, the singlet and triplet excitations are both gapless at $(0,0)$, $(\pi,0)$, $(0,\pi)$, and $(\pi,\pi)$.

To confirm this picture, we show in Fig.~\ref{fig_excitation_spectrum} the results for (a) singlet and (b) triplet excitation energies for $8\times8$, $12\times12$, and $16\times16$ lattices at $J_2 =0.5$ in the QSL phase. 
We compute not only at high-symmetry ${\bf K}$ points $(0,0)$, $(\pi,0)$, and $(\pi,\pi)$ but also at intermediate points $(\pi/2,0)$, $(\pi,\pi/2)$, and $(\pi/2,\pi/2)$ 
[and symmetrically equivalent ${\bf K}$ points such as $(-\pi/2,0)$, $(0,\pi/2)$, $(0,-\pi/2)$ for $(\pi/2,0)$].

We find that the excitation gap decreases as $L$ increases at the high-symmetry ${\bf K}$ points.
The exceptional behavior at ${\bf K} = (0,0)$ in the singlet sector is presumably an artifact, which arises from numerical difficulty in obtaining excited states in $S=0$ and ${\bf K} = (0,0)$ sector (Supplementary Note 2 in Appendix~\ref{Appendix_Notes}).
On the other hand, the gap stays nearly constant at the intermediate ${\bf K}$ points. 
By combining the gap analysis at the critical points (see above) 
and the size extrapolation of the gap by the scaling $a+b/L$ at the intermediate ${\bf K}$ points, 
we draw dispersion expected in the thermodynamic limit. 
The excitation spectra in the thermodynamic limit exhibit unconventional behavior in which the gap vanishes at the four high-symmetry momenta. 
We find only these four points as the gapless excitations suggesting Dirac-type linear dispersion around these four points.
To corroborate the conclusion about the four Dirac-type gapless points in the QSL phase, 
we have also calculated the excitation energies at $( m \pi/3,  n \pi/3)$ with $m,n = 0$, 1, 2, 3 for $12\times 12$ lattice (Fig.~\ref{fig_12x12_unit_cell_dep} in Appendix~\ref{Appendix_data}). 
From the limited momenta we studied, although other possibilities such as the higher-order dispersion (e.g., quadratic band touching) 
or tiny but extended gapless regions rather than points are not excluded, the results in Fig.~\ref{fig_12x12_unit_cell_dep} also support the Dirac-type nodal QSL.

\subsection{Signature of fractionalization in quantum spin liquid}
In the present QSL phase, one can expect an exotic fractionalization of particles, where a charge-neutral spin-1/2 excitation, called spinon, emerges. 
Although the spinon excitation cannot be detected experimentally, the evidence of the fractionalization can be detected as an incoherent continuum in the dynamic spin structure factor $S_{\rm s}({\bf q}, \omega)$ (spin-1 excitation)~\cite{Shao_2017}, which is interpreted by the two-particle (two-hole) or particle-hole excitation continuum of the spinons. 
We here compute the weight in $S_{\rm s}({\bf q}, \omega)$ at ${\bf q} = (\pi,0)$ and $(\pi,\pi)$ for the lowest triplet excitation shown in Fig.~\ref{fig_excitation_spectrum}. 
If the excitation were the conventional magnon branch of a magnetic phase, the weight would be the order 1.
If the weight vanishes, most of the weight lies in incoherent continuum at higher energies, supporting the emergence of fractionalized spinons~\cite{Shao_2017}.

Figure~\ref{fig_fractionalization}(a) shows the weight of the lowest branch in $S_{\rm s}({\bf q}, \omega)$ for ${\bf q} = (\pi,0)$ and $(\pi,\pi)$. 
We indeed see that the weight decreases as the system size increases. 
In particular, the weight at ${\bf q} = (\pi,0)$ rapidly decreases to zero, which means that the spectral weight is dominated by the incoherent continuum.
[We do not analyze the behavior at ${\bf q} = (\pi,\pi)$ in detail because of a numerical challenge due to the proximity to AF(N\'eel)-QSL phase boundary $J_2 = J_2^{\rm N\acute{e}el} \approx 0.49$ (Supplementary Note 3 in Appendix~\ref{Appendix_Notes})].
This is a strong evidence that the fractionalization indeed occurs in the QSL phase of the $J_1$-$J_2$ Heisenberg model. 
As we will discuss in Sec.~\ref{Sec_Discussion}, the dispersion of the emergent fractionalized spinon is expected to be gapless at the points $(\pm\pi/2,\pm\pi/2)$ [Fig.~\ref{fig_fractionalization}(b)].

\begin{figure}[tb]
\begin{center}
\includegraphics[width=0.46\textwidth]{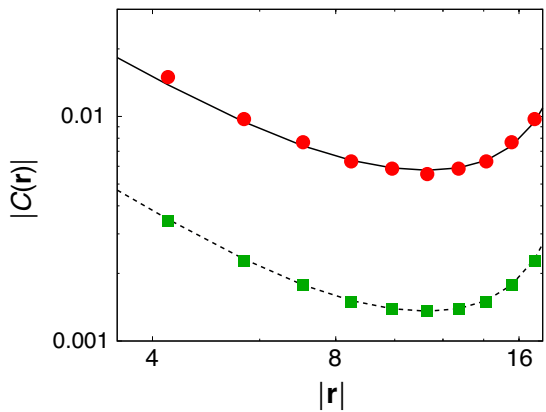}
\caption{
Real-space spin-spin (red dots) and dimer-dimer (green squares) correlation functions, $|C_{{ \rm s}} ( {\bf r})|$ and  $|C_{{ \rm d}_x} ( {\bf r})|$, respectively, for the diagonal direction ($r_x \! = \! r_y$) for $16\times16$ lattice at $J_2 = 0.5125$ in the QSL phase. 
 The solid and dashed lines are proportional to the power-law decay 
 $C ({\bf r})  \propto   \frac{1}{| {\bf r } |^{z+\eta}}  + \sum_{{\bf n} \neq (0,0) }    \left(   \frac{1}{|{ \bf r} - L {\bf n} |^{z+\eta}}  -  \frac{1}{|L {\bf n}|^{z+\eta}}   \right) $
 with $z+\eta = 1.52$ (solid) and 1.62 (dashed), in which we consider the effect of the periodic boundary condition~\cite{Morita_2015}. 
 The values of  $z+\eta$ are taken from the analysis in Fig.~\ref{fig_order_parameter}(c). 
 The upturn at large $| {\bf r } |$ is due to the periodicity of the lattice.  
 }
\label{fig_chi_vs_r}
\end{center}
\end{figure}

\subsection{Real-space correlation functions in the quantum spin liquid phase}
\label{subsec_real_space_correlation}
Figure.~\ref{fig_chi_vs_r} shows the real-space decay of spin-spin and dimer-dimer correlation functions, $|C_{{ \rm s}} ( {\bf r})|$ and  $|C_{{ \rm d}_x} ( {\bf r})|$, respectively, for the diagonal direction ($r_x \! = \! r_y$) for $16\times16$ lattice in the QSL phase ($J_2 = 0.5125$) (for the definition of the correlation function, see Methods). 
If the correlation function shows power-law decay, it is expressed by the exponent $z+\eta$, namely the spin-spin and dimer-dimer correlations should show $C({\bf r}) \sim r^{-(z + \eta)}$ ($r = |{\bf r}|$) in the real space as critical behavior.
Both correlation functions indeed show consistent behaviors with the power-law decay.
It evidences the dual critical nature of the VBS and N\'eel-AF correlations in the QSL phase, and this ground-state property is consistent with the gapless singlet and triplet excitations clarified independently (Sec.~\ref{sec_results_gapless}).
On top of the one-to-one correspondence between the ground-state phases and excitation structures revealed by the correlation ratio and level spectroscopy analyses (Secs.~\ref{sec_results_correlation_ratio} and \ref{sec_results_level_spectroscopy}), we again demonstrate a nice correspondence between the ground-state and excitation properties. 

\begin{figure}[tb]
\begin{center}
\includegraphics[width=0.48\textwidth]{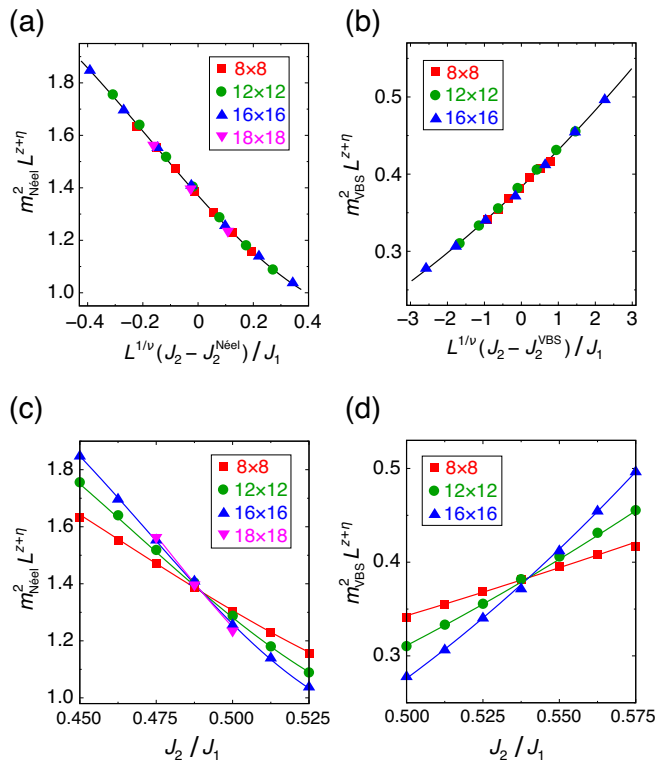}
\caption{
Finite-size scaling analysis. 
(a,c) Data collapse for N\`eel-AF order parameter.
We assume $J_2^{\rm N\acute{e}el} /J_1 =0.49$ and estimate the critical exponents $z+\eta$ and $\nu$. 
The Bayesian scaling analysis~\cite{Harada_2011,Harada_2015} gives $z+\eta = 1.410 (4)$ and  $\nu = 1.21 (5)$. 
(b,d) Data collapse for the VBS order parameter.
The same analysis with assuming $J_2^{\rm VBS} /J_1 =0.54$ gives $z+\eta = 1.436 (6)$ and  $\nu = 0.67(2)$.
Solid curves are the inferred scaling functions.
In (a) and (c), the $18\times18$ data are added to corroborate the result.
The figure shows that the conventional finite-size scaling analysis consistently supports the results obtained by the correlation ratio and level spectroscopy.
}
\label{fig_finite_size_scaling}
\end{center}
\end{figure}

\begin{figure*}[tb]
\begin{center}
\includegraphics[width=0.92\textwidth]{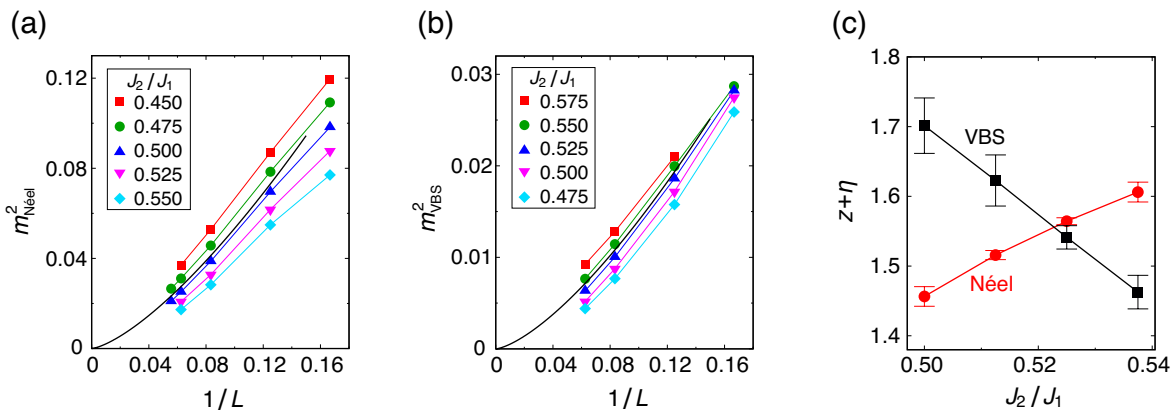}
\caption{
Size dependence of the squared order parameters for (a) N\`eel-AF and (b) VBS. 
The solid black curves in (a, b) are the expected size dependence at the critical point $m^2 \propto L^{-(z+\eta)}$ 
with $z+\eta$ estimated 
by the finite-size scaling analysis shown in Fig.~\ref{fig_finite_size_scaling}. 
The critical points are estimated from the analyses based on the correlation ratio and the level spectroscopy. 
(c) $J_{2}$ dependence of the power-law exponent $z+\eta$ in the QSL phase obtained by fitting the size dependence of $m^2$ for $L=8$, 12, 16 with a form $m^2 = A L^{-(z+\eta)}$ ($A$: constant).
}
\label{fig_order_parameter}
\end{center}
\end{figure*}

\subsection{Finite-size scaling and size dependence of order parameter}
Figures~\ref{fig_finite_size_scaling}(a) and \ref{fig_finite_size_scaling}(b) show the data of finite-size scaling analysis of the N\'eel-AF and VBS order parameters, respectively.
The squared order parameters for N\'eel-AF and VBS are given by $m_{\rm N\acute{e}el}^2 =   S_{\rm s} ({\bf Q}) / N_{\rm site} $ with ${\bf Q} = (\pi,\pi)$ and $m_{\rm VBS}^2 = S_{{\rm d}_x} ({\bf Q}) / N_{\rm site}$ with ${\bf Q} = (\pi,0)$ [$=  S_{{\rm d}_y} ({\bf Q}) / N_{\rm site}$ with ${\bf Q} = (0,\pi)$], respectively 
[see Methods for the finite-scaling analysis method and the definition of the structure factor, $S_{\rm s} ({\bf Q})$  and $S_{{\rm d}_x} ({\bf Q})$]. 
For the N\'eel-AF and VBS orderings, we assume that the critical points are at $J_2^{\rm N\acute{e}el} = 0.49$ and $J_2^{\rm VBS} =0.54$, respectively (see the phase diagram in Fig.~\ref{fig_phase_diagram}). 
The estimated critical exponents $z+\eta$ and $\nu$ deduced from the finite-size scaling are $z+\eta = 1.410 (4)$ and  $\nu = 1.21 (5)$ for the N\'eel-AF order parameter, and $z+\eta = 1.436 (6)$ and  $\nu = 0.67(2)$ for the VBS order parameter, respectively [The estimate does not depend significantly on the values of $J_2^{\rm N\acute{e}el}$ and $J_2^{\rm VBS}$ (Supplementary Note 4 in Appendix~\ref{Appendix_Notes})]. 
These exponents do not belong to the known universality class and suggest unconventional criticality.

Figs.~\ref{fig_order_parameter}(a) and \ref{fig_order_parameter}(b) show the size dependence of the N\'eel-AF  and VBS order parameters, respectively. 
Solid black curves are expected scaling curve $m^2 \sim L^{-(z+\eta)}$ at the critical points obtained by employing $z+\eta=1.410$ and 1.436 for the N\'eel-AF and VBS critical points, respectively.


\section{Discussion}
\label{Sec_Discussion}

As is discussed in Sec.~\ref{sec_J1J2model}, to settle the highly controversial situation on the phase diagram, the calculations need to fulfill three conditions: 
1. systematic $J_2$ dependence survey, 2. high accuracy, and 3. reliable estimate of the thermodynamic limit.
The RBM+PP data achieves the state-of-the-art accuracy level both for ground-state and excited states, satisfying the condition 2. 
With the high accuracy, we have performed a systematic investigation on the $J_2$ dependence both for ground-state and excited-state properties (condition 1). 
For the condition 3, both the correlation ratio and level spectroscopy have given consistent results, supporting the conclusion of the QSL phase in the region $0.49\lesssim J_2/J_1\lesssim 0.54$.
We do not find such quantitative consistency before, and we became convinced of the existence of the QSL phase only after finding their consistency. 
Nevertheless, we note that, at a qualitative level, an overall consensus on the existence of the QSL is being formed among the best accurate methods (Refs.~\onlinecite{hu13} and \onlinecite{gong14} and ours) clarified in the benchmark shown in Appendix~\ref{Appendix_Benchmark} (Note that Ref.~\cite{gong14} obtained essentially vanishing order consistent with our finite QSL region, though they considered alternative possibilities as well, which was not settled within their analyses of the size dependence of the order parameter correlation).



The spin excitation dispersion has been rarely studied in the literature except for the studies obtained by assuming {\it a priori} a variational form of $Z_2$ nodal spin-liquid wave function~\cite{Ferrari_2018,PhysRevB.98.134410}.
In Ref.~\onlinecite{PhysRevB.98.134410}, the spin cluster perturbation method is also employed to draw the dispersion. 
Our gapless structure lends support to these variational and the spin cluster perturbation studies in qualitative features, though our results have been obtained without such assumptions and approximations.    
Together with the consideration on the stability of the QSL phase~\cite{PhysRevB.65.165113} and the reason discussed below,
our unbiased analysis evidences the QSL phase in the $J_1$-$J_2$ Heisenberg model characterized by $Z_2$ nodal QSL (rather than $U(1)$ QSL) with gapless and fractionalized spin-$1/2$ spinon excitations at $(\pm \pi/2 ,\pm \pi/2)$, proposed in an earlier study~\cite{hu13} (we did not exclude the possibility of $U(1)$ QSL just from the spin excitation spectra because the $Z_2$ and $U(1)$ QSL give very similar $S_{\rm s}({\bf q}, \omega)$~\cite{PhysRevB.65.165113}).

The real spin excitations measurable in experiments must be made of two-spinon excitations, and thus the singlet and triplet gapless points are $(0,0)$, $(\pi,0)$, $(0,\pi)$ and $(\pi,\pi)$ [Fig.~\ref{fig_fractionalization}(c)]. 
The gapless Dirac-type excitations in both singlet and triplet sectors show an excellent consistency with the dual critical nature of the VBS and N\'eel-AF correlations, decaying algebraically in the real space, in the QSL phase (Sec.~\ref{subsec_real_space_correlation}).

The finite-size scaling analysis shown above suggests that the value for critical exponent $z + \eta$ is about 1.4 for both of the AF-QSL and QSL-VBS critical points (Fig.~\ref{fig_finite_size_scaling}), 
which is suggestive of an emergent symmetry between the spin-spin and dimer-dimer correlations, associated with the N\'eel-AF and VBS orders, respectively.
If the $U(1)$ QSL is realized as a phase, we will see the emergent symmetry within the whole QSL region as the critical phase~\cite{Hermele_2005,Xu_2019}. 
However, the power-law exponent $z+\eta$ seems to change in the QSL region: 
While it increases as $J_2$ increases for the spin-spin correlation, the dimer-dimer correlation shows the opposite trend [Fig.~\ref{fig_order_parameter}(c)].
It supports that the QSL with the emergent $U(1)$ symmetry is absent for an extended $J_2$ region and implies the extended region of the $Z_2$ QSL instead.
From Fig.~\ref{fig_order_parameter}(c), $U(1)$ symmetry is deduced to emerge at a single point $J_2^{U(1)} \approx 0.52$, where the values of $z+\eta$ for the spin-spin and dimer-dimer correlations cross and coincide, and the $Z_2$ QSL may have different characters between $J_2>J_2^{U(1)}$ and $J_2<J_2^{U(1)}$.
It will be of great interest to investigate this issue further in the future, especially by considering more detailed system size dependence to further establish the thermodynamic behavior. 

Since the excitation structure is isomorphic with the charge and spin excitations of the $d$-wave superconducting state in the cuprate superconductors, it is suggestive of the connection of the two; the superconducting state could be borne out from the present QSL immediately when carriers are doped. 
The present accurate estimate of the spinon excitation, especially, incoherent nature of the spin excitations with continuum, will provide us with insights into the unsolved puzzles of the cuprate superconductors including the incoherent transport and charge dynamics.

\section{Summary}

We have studied the 2D $J_1$-$J_2$ Heisenberg model using a highly accurate machine-learning method, RBM+PP.
Our achievements are summarized into the following points: 
the quantitative estimate of the phase diagram, useful insights into the QSL property to understand its nature, and the establishment of one-to-one correspondence between ground-state and excitation structure.

First, by combining the RBM+PP with the correlation ratio and level spectroscopy methods, we have been able to extrapolate to the thermodynamic limit reliably by two independent analyses. 
The agreement reached between the two at an unprecedented level has given the firm evidence for a finite QSL region $0.49\lesssim J_2/J_1\lesssim 0.54$.
The phase diagram is summarized in Fig.~\ref{fig_phase_diagram}.

The QSL is characterized as the dual nature of the algebraic and coexisting correlations of the antiferromagnetic (associated with the N\'eel order) and dimer (associated with the VBS order) correlations, which had been thought incompatible before by the symmetry difference. 
The elucidated dual nature is also seen consistently in the excitation property: 
We have identified the Dirac-type dispersion with gapless points $(0,0)$, $(\pi,0)$, $(0,\pi)$, and $(\pi,\pi)$ in both the singlet and triplet excitation sectors (related each to the dimer-dimer and spin-spin correlations, respectively).
The excitation structure is consistent with the emergence of the fractionalized spin-1/2 spinons with gapless Dirac dispersion. 
Interestingly, the power-law decay exponents of these two correlations change as a function of $J_2/J_1$ and do not coincide except for a single point around $J_2=0.52$, which imposes a substantial constraint on the gauge structure of the QSL. 

Finally, our comprehensive calculations have revealed a fundamental ``law of correspondence'' between the ground-state and excitation structure in the $J_1$-$J_2$ Heisenberg model. 
By establishing the phase diagram, we have demonstrated that the evolution of the ground state indeed maps to the change in the excitation structure induced by the level crossing in a fingerprint fashion with one-to-one correspondence. 
We have also shown that the coexisting power-law decay of the dimer-dimer and spin-spin correlation functions in the real space in the QSL phase (ground-state property) consistently corroborates the gapless structure of singlet and triplet excitations, respectively. 
Such one-to-one correspondence has a fundamental significance in physics, as the one-to-one correspondence between the equilibrium and non-equilibrium excited states addressed in the fluctuation-dissipation theorem and Kubo formula gives a foundation for the understanding of the linear response.

Such accurate, systematic, and comprehensive elucidation of the QSL with insights into the duality of the gapless correlations and the law of correspondence has been enabled by the RBM wave function combined with the PP state and the quantum number projection that offers state-of-the-art accuracy within a tractable computational cost: 
The high accuracy and the tractable computational-cost scaling of the RBM+PP method [${\mathcal O}(N_{\rm site}^3)$] were necessary to prepare comprehensive high-quality data for large system sizes to accomplish our achievement. 

So far, the machine learning methods had been applied mostly to benchmark problems with known solutions.  
By combining the RBM+PP wave function with cutting-edge methods to reduce finite-size corrections, we have succeeded in uncovering QSL in the long-standing challenging problem. 
This achievement opens a new avenue of numerical methods applicable to the grand challenges of quantum many-body systems.

\acknowledgements
We acknowledge useful discussions with Satoshi Morita, Anders W. Sandvik, and Zi Yang Meng. 
We also thank Satoshi Morita for providing us with the raw data in Ref.~\onlinecite{Morita_2015}.
Y.N. is grateful for fruitful discussions with Ribhu Kaul, Hidemaro Suwa, Yoshitomo Kamiya, Kenji Harada, Zheng-Cheng Gu, Giuseppe Carleo, and Ryui Kaneko.   
The implementation of the RBM+PP scheme is done based on the mVMC package~\cite{MISAWA2019447}.
The computation was mainly done at Supercomputer Center, Institute for Solid State Physics, University of Tokyo, and RIKEN supercomputers K and Fugaku.
The authors are grateful for the financial support by a Grant-in-Aid for Scientific Research (Grant No. 16H06345)  from Ministry of Education, Culture, Sports, Science and Technology (MEXT), Japan.
Y.N. was supported by Grant-in-Aids for Scientific Research (JSPS KAKENHI) (Grants No. 17K14336 and No. 18H01158).
This work is financially supported by the MEXT HPCI Strategic Programs, and the Creation
of New Functional Devices and High-Performance Materials to Support Next Generation Industries (CDMSI) as well as by ``Program for Promoting Researches on the Supercomputer Fugaku" (Basic Science for Emergence and Functionality in Quantum Matter
- Innovative Strongly-Correlated Electron Science by Integration of ``Fugaku" and Frontier Experiments -) (Grants No. hp200132 and hp210163).
We also acknowledge the support provided by the RIKEN
Advanced Institute for Computational Science under the
HPCI System Research project (Grants No. hp170263, hp180170, and hp190145).

\appendix

\setcounter{secnumdepth}{1}

\section{Methods --Detail}
\label{Appendix_method_detail}

\subsection{Optimization of RBM+PP wave function} 
\label{Appendix_optimization}

To search the lowest-energy quantum state for each quantum number sector, we optimize the variational parameters $\{ b_k, W_{ik},  f_{ij}^{\uparrow \downarrow} \}$ to minimize the energy expectation value of the RBM+PP wave function. 
The energy expectation value $ E =  \frac{ \langle \Psi | { \mathcal H }  | \Psi  \rangle  } { \langle \Psi | \Psi \rangle  }$ can be calculated by the Monte Carlo sampling with weight $p(\sigma)   =   \frac{ | \Psi (\sigma) |^2  }{\langle \Psi |\Psi  \rangle  }  $
\begin{eqnarray}
\label{Eq_tot_ene}
     E  =\sum_{\sigma}     p (\sigma)  E_{\rm loc} (\sigma), 
\end{eqnarray}
where the local energy $E_{\rm loc} (\sigma)$ is given by 
$E_{\rm loc} (\sigma) =  \sum_{\sigma'}  \langle \sigma |  { \mathcal H }   | \sigma ' \rangle  \frac{  \langle \sigma' |  \Psi  \rangle } { \langle \sigma |  \Psi \rangle }$.
The $E$ value depends on the variational parameters. 
To optimize the variational parameters to minimize $E$, we employ the stochastic reconfiguration (SR) method~\cite{PhysRevB.64.024512}, 
which is equivalent to the imaginary-time Hamiltonian evolution $e^{-\tau {\mathcal H}}  | \Psi  \rangle$  within the Hilbert space spanned by the RBM+PP wave function. 
Because the imaginary-time Hamiltonian evolution $e^{-\tau {\mathcal H}}  | \Psi  \rangle$  always stably gives the lowest-energy state for each quantum number sector (as far as the initial RBM+PP state is not orthogonal to the lowest-energy state), the SR method enables stable optimizations. 
For further technical details of the SR optimization, we refer to Ref.~\onlinecite{PhysRevB.96.205152}.

The number of complex variational parameters in the RBM part is $N_{\rm hidden} $  for $b_k$ and $N_{\rm hidden} \times N_{\rm site}$ for $W_{ik}$, respectively. 
As for the real variational parameters $f_{ij}^{\uparrow \downarrow}$ in the PP part,
to reduce the number of parameters and the computational cost, 
we impose $4 \times 4$ sublattice structure for the $8\times8$, $12\times12$, and $16\times16$ lattices,
and $6 \times 6$ sublattice structure for the $18\times18$ lattice, 
whereas we do not employ sublattice structure for the $6\times6$ lattice. 
In the case of the $4 \times 4$ sublattice structure, the number of independent $f_{ij}^{\uparrow \downarrow} $ parameters is reduced from $N_{\rm site}^2$ to $4 \times 4 \times N_{\rm site} = 16N_{\rm site}$, 
and the other $f_{ij}^{\uparrow \downarrow} $ parameters are defined by spatial translation operations.  
In the presence of the Gutzwiller factor to map the PP state onto the spin system, 
the onsite $f_{ii}^{\uparrow \downarrow}$ parameters become completely redundant, i.e., the wave function does not depend on $f_{ii}^{\uparrow \downarrow}$  at all. 
Then, the number of relevant $f_{ij}^{\uparrow \downarrow}$ parameters is $16(N_{\rm site} \! - \! 1)$.
For the initial values for $\{ b_k, W_{ik},  f_{ij}^{\uparrow \downarrow} \}$, we put random numbers in order not to introduce bias in the initial variational state. 
More specifically, for each real and complex part of $b_k$ and $W_{ik}$ parameters, we put small random numbers from the interval $[-0.05,0.05]$.
In the case of the triplet state calculation, $b_k$ parameters are multiplied by 10 (note that if $b_k$ is zero, the RBM part is completely symmetric with respect to the global spin inversion).
For the initial $f_{ij}^{\uparrow \downarrow}$ parameters, we put random numbers from $[-f_{\rm max}, f_{\rm max}]$ with $f_{\rm max}$ depending on the distance $R_{ij}$ between $i$th and $j$th sites. 
We typically take $f_{\rm max}$ to be proportional to $R_{ij}^{-a}$ with $a \sim 2$.
Random $f_{ij}^{\uparrow \downarrow}$ parameters allow various spin ordering patterns whose period is within the sublattice size. 
A longer period structure than the system size is beyond the scope of this study, as are all the other earlier finite-system-size studies.
For each $J_2$ point, we perform at least three independent optimizations of the RBM+PP wave functions from different initial variational parameters.  
We discuss the initial-parameter dependence in more detail in Appendix~\ref{Appendix_initial_guess_dep}.

The computational cost of the RBM+PP wave function employing sublattice structure in the PP part scales with ${\mathcal O}( N_{\rm site}^3)$.
In the RBM+PP method, a computationally demanding part is coming from the calculation of the PP wave function part
and the neural network (RBM) part offers an efficient way of improving accuracy.

\subsection{Special treatments to obtain some specific excited states}   
\label{Appendix_excited}

As we describe in Sec.~\ref{sec_qp_proj}, we apply the spin-parity projection to distinguish whether the total spin $S$ is even or odd. 
Because the singlet (triplet) state is the lowest state for each even (odd) $S$ sector in the present study,  
we obtain a singlet (triplet) state for the even (odd) $S$ sector. 
Therefore, we can obtain the singlet ($S=0$) and triplet ($S=1$) excited states with momentum resolution. 
However, we need special treatment to obtain $S=0$ excited state at ${\bf K} = (0,0)$ 
because the lowest-energy state in $S=0$ and ${\bf K} = (0,0)$ quantum number sector is the ground state. 
We use additional simplified point-group projection on top of those in Eq. (\ref{Eq.quantum_proj}) to obtain excited states belonging to a different irreducible representation of the $C_{4v}$ point group of the square lattice than that of the ground state as follows: 
\begin{eqnarray}
 \Psi_{{\bf K} =  (0,0)}^{A, S_+} ( \sigma )  &=&  
    \Psi_{{\bf K} =  (0,0)}^{S_+} ( \sigma)  +  \Psi_{{\bf K} =  (0,0)}^{S_+} ( R_{\pi/2} \sigma)
    \label{Eq_A_projection} \\ 
   \Psi_{{\bf K} =  (0,0)}^{B, S_+} ( \sigma )  &=&  
    \Psi_{{\bf K} =  (0,0)}^{S_+} ( \sigma)  -  \Psi_{{\bf K} =  (0,0)}^{S_+} ( R_{\pi/2} \sigma), 
\end{eqnarray}
where the $R_{\pi/2}$ is an operator to rotate the spin configuration by 90 degrees. 
With this projection, we can distinguish whether the state belongs to $A$ (either $A_1$ or $A_2$) irreducible representation or $B$  (either $B_1$ or $B_2$) irreducible representation under the $C_{4v}$ point group (to distinguish between $A_1$ and $A_2$ or between $B_1$ and $B_2$, we need full point group projection with $0$, $\pi/2$, $\pi$, $3\pi/2$ rotations).
The ground state corresponds to the former, while the excited state corresponds to the latter.

We also need special treatment to obtain $S=2$ excited states. 
To this end, we use the mVMC (many-variable variational Monte Carlo method)~\cite{MISAWA2019447} based on the PP wave function. 
In the mVMC, the full spin projection to specify the total spin is available, and we apply it to get $S=2$ states. 
The full spin projection is time-consuming (at least about five times) compared to the spin-parity projection.  
At the cost of longer computational time for the full spin projection, the mVMC (only PP) gives comparable accuracy to the RBM+PP method.

\begin{figure*}[tb]
\begin{center}
\includegraphics[width=0.95\textwidth]{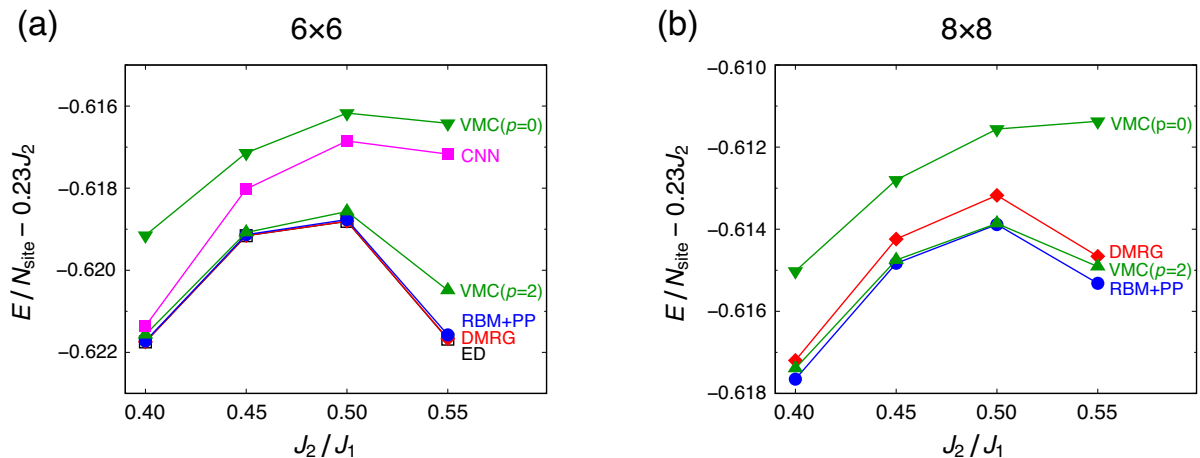}
\caption{
Comparison of the ground state energy for the $J_1$-$J_2$ Heisenberg model. 
The comparison is made among the variational energies under the periodic boundary condition.  
The system sizes are (a)  $6\times6$ and (b)  $8\times8$. 
Our RBM+PP results are compared with those obtained by the variational Monte Carlo (VMC) method combined with the $p$-th order Lanczos steps~\cite{hu13}, 
the density-matrix renormalization group (DMRG) (with 8182 $SU$(2) states)~\cite{gong14}, the convolutional neural network (CNN)~\cite{Choo_2019}, and the exact diagonalization (ED)~\cite{Schulz_1996}. 
The CNN and ED results are available only for the $6\times6$ lattice. 
}
\label{fig_ene_comparison}
\end{center}
\end{figure*}

\subsection{Calculation conditions} 

In the present study, we fix the number of hidden units $N_{\rm hidden}$ to be $16$. 
We always apply the spin-parity and momentum projections during the optimization of the RBM+PP wave function. 
The special treatments to obtain $S=0$ excited state at ${\bf K} = (0,0)$ and $S=2$ excited states are described above. 
To improve the quality of the data for the correlation function in Figs.~\ref{fig_ratio_crossing}, \ref{fig_fractionalization}, \ref{fig_chi_vs_r}, \ref{fig_finite_size_scaling}, \ref{fig_order_parameter}, \ref{fig_AF_structure_factor}, \ref{fig_VBS_structure_factor}, and \ref{fig_ratio_cross_size_dep} quantitatively, 
we apply the simplified point-group projection in Eq. (\ref{Eq_A_projection}) to the optimized ground state RBM+PP wave function for the sector with $S=0$ and ${\bf K} = (0,0)$. 
The ground state energy in Fig.~\ref{fig_ene_comparison} is also produced with the simplified point-group projection (see Appendix~\ref{Appendix_Benchmark_Accuracy} for the details).

\section{Benchmark}
\label{Appendix_Benchmark}

\subsection{Accuracy of the RBM+PP wave function}
\label{Appendix_Benchmark_Accuracy}

By applying the RBM+PP method to the 2D $J_1$-$J_2$ Heisenberg model on the square lattice, we confirm that the RBM+PP 
achieves state-of-the-art accuracy 
 not only among machine-learning-based methods~\cite{PhysRevB.98.104426,Choo_2019,Ferrari_2019,Westerhout_2020}
but also among all available numerical methods. 
Figure~\ref{fig_ene_comparison} shows the comparison of the ground-state energy among various methods for the $6\times 6$ and $8\times 8$ lattices (see Table~\ref{table:GS_ene} for the raw data). 
Here, the RBM+PP energy is obtained by optimizing the RBM+PP wave function with the momentum, spin-parity, and simplified-point-group projections. 
We do not employ the sublattice structure in the $f_{ij}^{\uparrow \downarrow}$ parameters (the sublattice structure used in the actual calculations is discussed in Appendix~\ref{Appendix_optimization}), and the number of hidden units is 16 as commonly employed in the paper.  
Up to the $6\times 6$ lattice, the exact diagonalization result is available. 
At $J_2 = 0.5$, where the frustration is strong, the relative error of the RBM+PP energy is less than 0.01 \%, demonstrating the high accuracy of the RBM+PP wave function. 
For the $8 \times 8$ lattice, the RBM+PP wave function gives the best accurate energy among the compared variational methods for all $J_2$ values we studied.

We have also performed the benchmark calculations for the $10 \times 10$ lattice because the benchmarks of neural-network wave functions in the literature have mainly been performed using the $10 \times 10$ lattice. 
We optimized the RBM+PP wave function with 16 hidden units (as used in the other system sizes) without introducing a sublattice structure in the PP part. 
We apply the momentum, spin-parity, and point-group projections.
Table~\ref{table:ene_comparion_10x10} shows the comparison of the ground-state energy at $J_2=0.5$ among different wave functions. 
As in the $8 \times 8$ lattice result, the RBM+PP gives the best accuracy among the various methods. 
From the systematic benchmarks on the $6\times6$, $8\times8$, and $10\times10$ lattices, we conclude that the RBM+PP achieves state-of-the-art accuracy.

\begin{table}[tb]
\caption{
Raw data of RBM+PP ground-state energy in Fig.~\ref{fig_ene_comparison}.  
}
\label{table:GS_ene}
\centering 
\vspace{0.3cm}
\begin{tabular}{ccccc}
\hline \hline
            &  $J_2 = 0.40$ & $J_2 = 0.45$  & $J_2 = 0.50$  & $J_2 = 0.55$  \\ \hline
$6\!\times\!6$ & $-0.529726$(1)  & $-0.515633$(1)  & $-0.503765$(1) & $-0.495075$(1) \\
$8\!\times\!8$ & $-0.525653$(1)  & $-0.511331$(1)  & $-0.498886$(1) & $-0.488820$(2) \\ \hline \hline
\end{tabular} 
\end{table}

\begin{table}[tb]
\caption{
Comparison of ground-state energy for the $10\times 10$ lattice at $J_2=0.5$ among different wave functions.
The wave functions in bold font use neural networks.
In Ref.~\cite{hu13}, $p$-th order Lanczos steps are applied to the VMC wave function.
 }
\label{table:ene_comparion_10x10}
\centering 
\vspace{0.3cm}
\begin{tabular}{@{\   }  l @{\  \  \  \   }  l @{\   \ }  l  }
\hline \hline
 Energy per site  &  Wave function &  Reference    \\ \hline
$-0.494757(12)$  & {\bf Neural quantum state}  &  \   \onlinecite{Szabo_2020} \\
$-0.49516(1)$    &  {\bf CNN}  &  \  \onlinecite{Choo_2019} \\
$-0.49521(1)$    &  VMC($p$=0) &  \  \onlinecite{hu13} \\
$-0.495530$       &  DMRG & \   \onlinecite{gong14} \\ 
$-0.49575(3)$    &  {\bf RBM-fermionic w.f.}&\   \onlinecite{Ferrari_2019} \\ 
$-0.497549(2)$  &  VMC($p$=2)  &\   \onlinecite{hu13} \\ 
$-0.497629(1)$  &  {\bf RBM+PP} & \  present study \\ 
\hline \hline
\end{tabular} 
\end{table}

In the actual calculations, we employ the sublattice structure in the $f_{ij}^{\uparrow \downarrow}$ parameters for the $8\times8$, $12\times12$, $16\times16$, and $18\times18$ lattices,
to reduce the computational cost from ${\mathcal O}( N_{\rm site}^4)$ to ${\mathcal O}( N_{\rm site}^3)$ (see Appendix~\ref{Appendix_optimization}).
The reduction of the computational time enables us to perform systematic calculations for various $J_2$ values and for different quantum-number sectors. 
By employing the sublattice structure, the accuracy becomes slightly worse compared to that without sublattice structure. 
For example, in the case of the $8\times8$ lattice at $J_2=0.5$, the ground-state energy with the $4\times4$ sublattice structure is 
$-0.498460$(6), which is compared to $-0.498886$(1) obtained without a sublattice structure. 
The difference is less than 0.1 \%; therefore, high accuracy is retained even with the sublattice structure\footnote{
For large system sizes, we employ sublattice structures to make the calculations ``practical'' (making computational cost manageable). 
For example, in Ref. \cite{hu13}, the VMC($p$=2) results are not available for large system sizes, and the ``practical'' calculation is VMC($p$=1).
We notice that, thanks to the retained accuracy, our calculations also show state-of-the-art accuracy at the ``practical'' level;  
for the ground-state energy for the $18 \times 18$ lattice at $J_2 = 0.5$, the VMC($p$=1) wave function in Ref.~\cite{hu13} gives $E/N_{\rm site} = - 0.49611(1)$, whereas our RBM+PP wave function with $6\times6$ sublattice structure gives a better precision of $E/N_{\rm site} = -0.496275(3)$
}.  
As for the spin-spin and dimer-dimer correlations, the obtained values of the order parameters are 
$m_{\rm N\acute{e}el}^2 = 0.06955(8)$ and  $m_{\rm VBS}^2 = 0.01720(3)$ in the case of the $4\times4$ sublattice structure, 
and  
$m_{\rm N\acute{e}el}^2 = 0.06724(8)$ and  $m_{\rm VBS}^2 = 0.01703(3)$ in the case of no sublattice structure.
The actual calculations with sublattice structure tend to slightly overestimate the order parameters in the frustrated regime. 
We see a similar tendency in the case of the benchmark calculations of RBM only wave functions for the $6 \times 6$ lattice at $J_2=0.5$ with changing the number of hidden units~\cite{Nomura_2021}, 
where the N\'eel-AF order parameter tends to be overestimated for a small number of hidden units.
With increasing the number of hidden units, the accuracy improves, and the order parameter shows an excellent agreement with the exact results~\cite{Nomura_2021}.
Considering the fact that improving accuracy tends to suppress the order parameter, 
our statement of the existence of the QSL phase with vanishing order parameters in the thermodynamic limit should be valid.

\begin{figure}[tb]
\begin{center}
\includegraphics[width=0.46\textwidth]{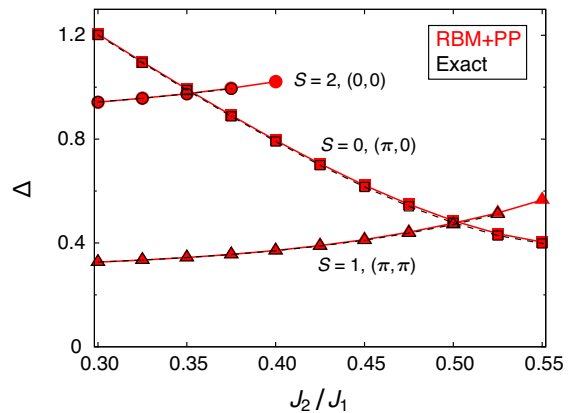}
\caption{
Singlet, triplet and quintuplet excitation energies for $6\times 6$ lattice obtained by RBM+PP (filled symbols) and ED (open symbols). 
In ED, we calculate up to five excited states using ${\mathcal H}\Phi$~\cite{KAWAMURA2017180}. 
At $J_2 = 0.40$, $S=2$ excitation with the momentum $(0,0)$ is not included in the five lowest excited states.  
The same holds for $S=1$ excitation with the momentum $(\pi,\pi)$ at $J_2 = 0.55$. 
The RBM+PP and exact results show a good agreement.
}
\label{fig_ex_ene_6x6}
\end{center}
\end{figure}

Remarkably, we also find that the RBM+PP accurately represents excited states as well as the ground state. 
Figure~\ref{fig_ex_ene_6x6} shows the comparison of excitation energies for singlet, triplet, and quintuplet excitations between the exact and RBM+PP results for the $6\times 6$ lattice. 
The agreement is excellent, where the difference in energy between the exact and the RBM+PP results is less than 0.01. 
Previously, there have been several attempts to obtain the excitation gap of the $J_1$-$J_2$ model~\cite{hu13,jiang12,gong14,Wang_2018}. 
In Ref.~\onlinecite{hu13} using the combination of the VMC and Lanczos methods, the excited states are obtained by changing boundary condition, which limits the number of excited states that can be calculated [only $S=2$ with the momentum $(0,0)$ and $S=0$ with $(\pi,0)$ or $(0,\pi)$].
Also, the accuracy does not reach the level shown in Fig.~\ref{fig_ex_ene_6x6} even with the 2nd-order Lanczos being applied [VMC($p=2$)].
In Refs.~\onlinecite{jiang12,gong14,Wang_2018} using the density-matrix renormalization group (DMRG), the open boundary condition is employed, and hence the dispersion is not available because the momentum is ill-defined.   
In the present study, we can obtain accurate excitation energies with momentum resolution.  
The accurate estimate of excitation gaps enables us to perform the level spectroscopy to estimate the phase boundary and elucidate the nature of the QSL phase.

\begin{figure}[tb]
\begin{center}
\includegraphics[width=0.48\textwidth]{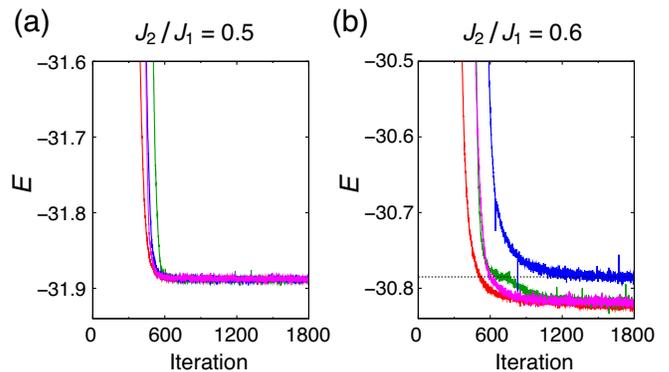}
\caption{
Initial-parameter dependence of the RBM+PP optimization curves for the ground state at (a) $J_2=0.5$ and (b) $J_2=0.6$ for the $8\times8$ lattice.
The results for the four independent optimizations are shown for each $J_2$. 
In (b), the dotted line indicates the total energy of a local-minimum solution.
}
\label{fig_initial_guess_dep}
\end{center}
\end{figure}

\subsection{Initial-parameter dependence of the RBM+PP optimization}
\label{Appendix_initial_guess_dep}

As we describe in Appendix~\ref{Appendix_optimization}, we perform several independent optimizations of the RBM+PP wave functions for each $J_2$ point. 
Here, using the $8\times 8$ lattice, we show how the difference in initial variational parameters affects the optimization. 
Figure~\ref{fig_initial_guess_dep} shows the initial-parameter dependence of the RBM+PP ground-state optimization curves for $J_2= 0.5$ and $J_2=0.6$. 
For $J_2$ = 0.5, we see that four independent optimizations converge to the same energy stably. 
On the other hand, at $J_2 = 0.6$, the RBM+PP wave function whose optimization curve is shown in blue color seems to be trapped in a local minimum. 
The green curve is also trapped at similar energy (dotted line), but it eventually gets out of the local minimum.
The behavior seen at $J_2 = 0.6$ can be understood from the proximity to the 1st-order transition point around $J_2 = 0.61$ between the VBS and stripe-AF phases.
At large system sizes, there exists an energy-level crossing between the states belonging to the same quantum-number sector (zero total momentum and singlet), which gives a kink in the $J_2$ dependence of the ground state energy (Fig.~\ref{fig_GS_ene}).
Therefore, different solutions are competing in small energy scale in the same quantum-number sector at $J_2 = 0.6$, which makes the optimization more unstable as compared to that at $J_2 = 0.5$. 

One of the reasons for the stable optimization is that finite-size systems we have treated [$N_{\rm site}$ is ${\mathcal O}(100)$] have a finite energy level spacing except for level crossing points. 
The order of the energy level spacing is on the order of 0.1 (see, e.g., Fig.~\ref{fig_level_cross_1}), and the RBM+PP method has a finer energy resolution (note that the energy axis scale of Fig.~\ref{fig_initial_guess_dep} is 0.1). 
The results in Fig.~\ref{fig_initial_guess_dep} suggest that, though the optimized variational parameters may depend on the initial parameters, the optimized wave functions themselves are essentially identical (we have confirmed this by calculating the overlap using the Monte Carlo method among the optimized wave functions).

From this benchmark, we notice that it is important to perform several independent optimizations to avoid being trapped in local minima. 
In the present study, although the optimizations of the RBM+PP wave functions are done independently for different $J_2$ points,
thanks to the several independent optimizations at each $J_2$ value, all the physical quantities change smoothly and continuously.

\section{Supplementary data}  
\label{Appendix_data}

\subsection{Ground state energy }

\begin{figure}[tb]
\begin{center}
\includegraphics[width=0.46\textwidth]{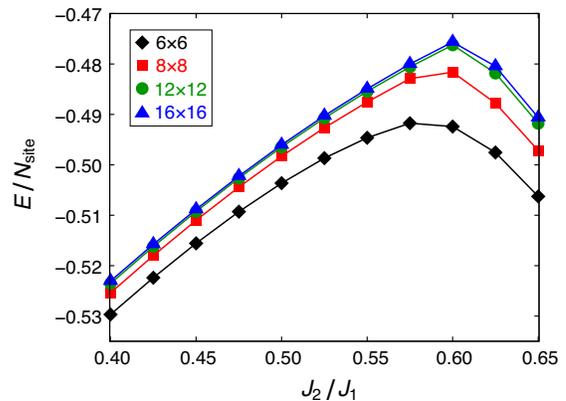}
\caption{
$J_2$ dependence of RBM+PP ground-state energy of square-lattice $J_1$-$J_2$ Heisenberg model. }
\label{fig_GS_ene}
\end{center}
\end{figure}

The phase transition between the VBS and stripe-AF phases at $J_2^{\rm V\mathchar`-S}$ in Fig.~\ref{fig_phase_diagram} is of 1st order. 
To see this, we show the ground state energy as a function of $J_2$ in Fig.~\ref{fig_GS_ene}. 
As the system size increases, we see a clear kink in the energy curve at $J_2^{\rm V\mathchar`-S} \approx 0.61$, giving evidence for the 1st-order phase transition.

\begin{figure}[tb]
\begin{center}
\includegraphics[width=0.5\textwidth]{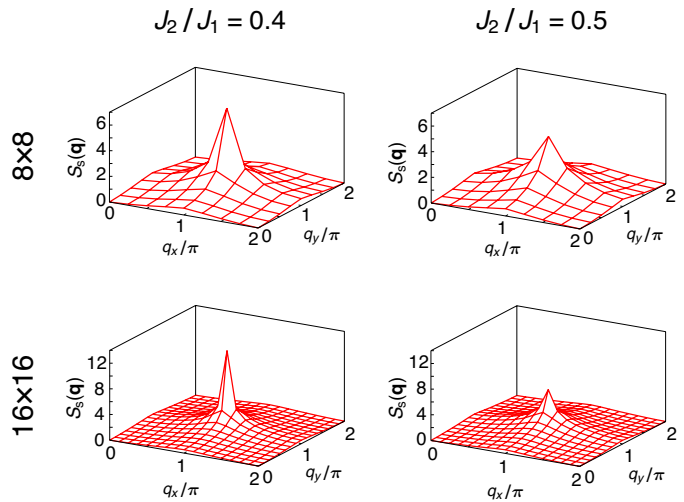}
\caption{
 Structure factor for spin-spin correlation $S_{\rm s}({\bf q})$. }
\label{fig_AF_structure_factor}
\end{center}
\end{figure}

\begin{figure}[tb]
\begin{center}
\includegraphics[width=0.5\textwidth]{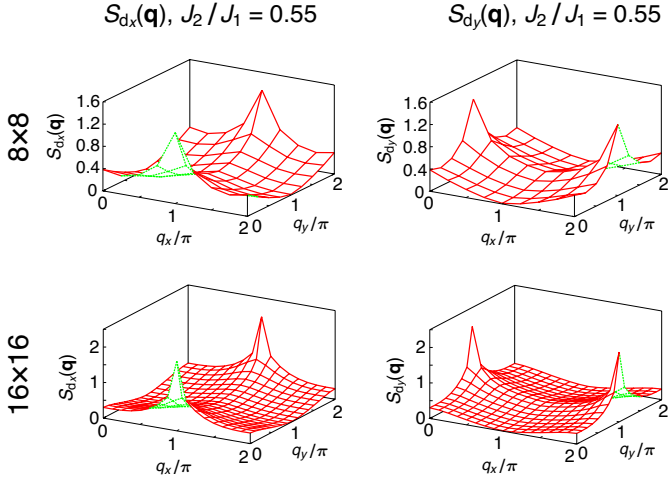}
\caption{
Structure factor for dimer-dimer correlation $S_{{\rm d}_x}({\bf q})$ and $S_{{\rm d}_y}({\bf q})$. }
\label{fig_VBS_structure_factor}
\end{center}
\end{figure}

\subsection{Structure factors}  

In Sec.~\ref{sec_results_correlation_ratio}, we discuss the crossing of the correlation ratio. 
The correlation ratio quantifies how sharp the structure factor peak is. 
In Figs.~\ref{fig_AF_structure_factor} and \ref{fig_VBS_structure_factor}, we show the raw data of the structure factors for spin-spin and dimer-dimer correlations, respectively, which are used in the correlation ratio analysis.

\subsection{System-size dependence of the crossing $J_2$ points of the AF and VBS correlation ratios}
\label{Appendix_ratio_size_dep}

\begin{figure}[tb]
\begin{center}
\includegraphics[width=0.46\textwidth]{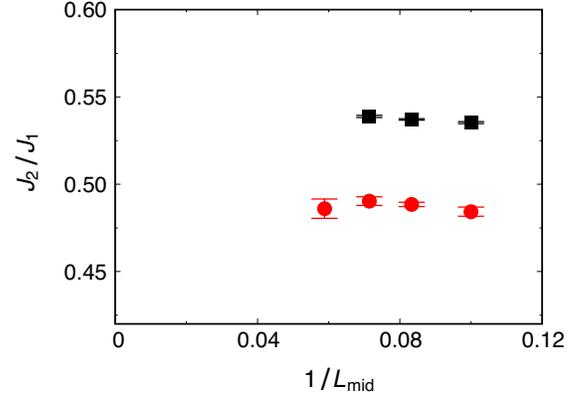}
\caption{
System-size dependence of the crossing points of the correlation ratio for spin-spin (red dots) and dimer-dimer (black squares) correlations, which are used to determine the phase boundary of N\'eel-AF and VBS, respectively.
We focus on the crossing of the curves between the $L_1\times L_1$ and $L_2 \times L_2$ lattices with $(L_1, L_2) = (8,12)$, $(8,16)$, $(12,16)$ and $(16,18)$ for spin-spin correlations, and $(L_1, L_2) = (8,12)$, $(8,16)$, and $(12,16)$ for the dimer-dimer correlations. 
$L_{\rm mid}$ is defined as $L_{\rm mid} = (L_1+L_2)/2$.
}
\label{fig_ratio_cross_size_dep}
\end{center}
\end{figure}

As described in Sec.~\ref{sec_results_correlation_ratio}, we determine the AF-QSL and QSL-VBS phase boundaries from the correlation ratio analysis.
Figure~\ref{fig_ratio_cross_size_dep} shows the system-size dependence of the crossing points of the correlation-ratio curves.
We see that the system-size dependence is small.
The fits of the system-size dependence with $a+b/L$ dependence give the estimates of AF-QSL and QSL-VBS phase boundaries as $J_2^{\rm N\acute{e}el} = 0.492(8)$ and $J_2^{\rm VBS} =0.548(1)$.
The fits using $a+b/L^2$ give $J_2^{\rm N\acute{e}el} = 0.490(4)$ and $J_2^{\rm VBS} =0.542(1)$.
These results support our conclusions of $J_2^{\rm N\acute{e}el}\approx 0.49$ and $J_2^{\rm VBS}\approx 0.54$. 

\subsection{Excitation gap at \mbox{\boldmath $12 \times 12$} lattice -- sublattice-size dependence in the PP part}

As we mentioned in Appendix~\ref{Appendix_optimization}, 
we impose the $4 \times 4$ sublattice structure in the $f_{ij}^{\uparrow \downarrow}$ parameters in the PP part. 
With this setting, we have momentum resolution of $4 \times 4$ ${\bf K}$ points: ${\bf K} = ( m \pi/2,  n \pi/ 2)$ with $m,n= -1$, 0, 1, 2. 
To investigate the sublattice-size dependence, for $12\times12$ lattice, 
we also calculate the excitation energies using $6 \times 6$ sublattice structure. 
Then, we can calculate the excitation gaps at ${\bf K} = ( m \pi/3,  n \pi/3)$ with $m,n = -2$, $-1$, 0, 1, 2, 3.

\begin{figure}[tb]
\begin{center}
\includegraphics[width=0.48\textwidth]{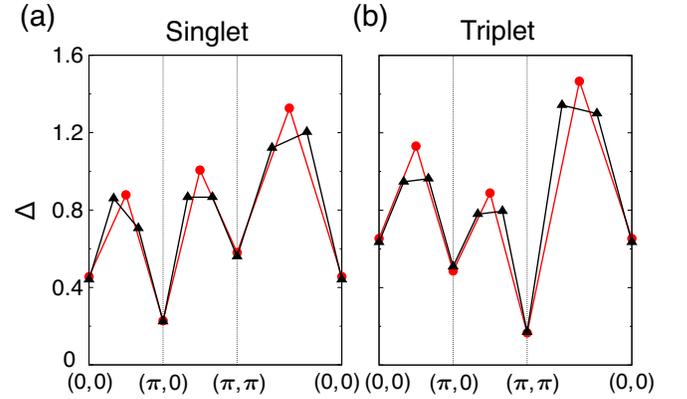}
\caption{
$f^{\uparrow \downarrow}_{ij}$-sublattice-size dependence of excitation.
(a) Singlet and (b) triplet excitation energy along the symmetric line in the Brillouin zone for $12\times 12$ lattice. 
Red dots: $4\times 4$ sublattice structure. 
Black triangles: $6\times 6$ sublattice structure. 
}
\label{fig_12x12_unit_cell_dep}
\end{center}
\end{figure}

Figure~\ref{fig_12x12_unit_cell_dep} shows the $f_{ij}^{\uparrow \downarrow}$-sublattice-size dependence of the excitation energies. 
We see that the excitation gaps at high-symmetry ${\bf K}$ points [$(0,0)$, $(\pi,0)$, and $(\pi,\pi)$] show good agreement between the  $4 \times 4$  and $6\times6$ sublattice structures.
At the intermediate ${\bf K}$ points, the excitation energies stay larger than those at high-symmetry ${\bf K}$ points. 
This fact supports the scenario of Dirac-type nodal QSL.

\section{Supplementary Notes}
\label{Appendix_Notes}

\begin{enumerate}

\item
Around the AF-QSL and QSL-VBS phase boundaries, we see noteworthy features in singlet excitations at ${\bf K}=(\pi,\pi)$ and triplet ones at ${\bf K}=(\pi,0), (0,\pi)$.
First, around the AF-QSL boundary ($J_2 = J_2^{\rm N\acute{e}el} \approx 0.49$), we see the kink in the excitation energy in the singlet ${\bf K}=(\pi,\pi)$ excitation [Figs.~\ref{fig_level_cross_1}(a) and \ref{fig_level_cross_1}(b)]. 
Actually, there is a level crossing in this quantum number sector, and the point-group irreducible representation of the lowest state changes at the kink. 
Also around the QSL-VBS boundary ($J_2 =J_2^{\rm VBS}\approx 0.54$), with increasing $J_2$, there is an upturn of the excitation energy of triplet ${\bf K}=(\pi,0)$ excitation for $16\times 16$ lattice 
[Fig.~\ref{fig_level_cross_1}(b)], which seems consistent with the fact that the triplet excitation has a gap in the VBS phase. 
These two supplementary features are suggestive of the connection to the phase transitions; it would be interesting to investigate them further.

\item
In Fig.~\ref{fig_excitation_spectrum}, the excitation energy with $S=0$ and ${\bf K} = (0,0)$ sector stays almost constant as the system size $L$ changes, in contrast with the behavior at the other high-symmetry ${\bf K}$ points. 
The singlet excited state at ${\bf K} = (0,0)$ must belong to a different irreducible representation than that of the ground state, because, in the present method, we cannot obtain the excited states with the same irreducible representation as that of the ground state. 
Such excited states with the same irreducible representation might show similar behavior to those at the other high-symmetry ${\bf K}$ points.

\item
The weight of the triplet at $(\pi,\pi)$ seems to be scaled naturally to a nonzero value, which might imply the remnant of the pole. 
This requires further clarification in larger system sizes in the future. 
The reason could partly be that the calculation is done close to the AF(N\'eel)-QSL phase boundary $J_2 = J_2^{\rm N\acute{e}el} \approx 0.49$. 
Another origin might be a possible anisotropic (elliptic) Dirac dispersion of spinons with preserved $C_4$ symmetry, which makes the spinon particle-hole excitation denser for the momentum transfer $(\pi,\pi)$ and makes the slow convergence to zero.

\item
The $J_2^{\rm N\acute{e}el}$ and $J_2^{\rm VBS}$ dependence of the estimate of the critical exponents is as follows.
For the N\'eel-AF order parameter, $z+\eta = 1.384(3)$, 1.410(4), 1.437(5) and $\nu =  1.22(4)$, 1.21(5), 1.18(5) for $J_2^{\rm N\acute{e}el} = 0.485$, 0.490, 0.495, respectively. 
The $\nu$ values for different  $J_2^{\rm N\acute{e}el}$ values agree within the size of error bars. 
Although $z+\eta$ increases as  $J_2^{\rm N\acute{e}el}$ increases, the values lie around 1.4. 

For the VBS order parameter, $z+\eta = 1.471 (8)$, 1.436(6), 1.400(5) and $\nu = 0.66(3)$,  0.67(2),  0.65(2) for $J_2^{\rm VBS} = 0.535$, 0.540, 0.545, respectively. 
Though $z+\eta$ decreases slightly as $J_2^{\rm VBS}$ increases, it lies between 1.4 and 1.5, which is close to those at the N\'eel-AF critical point.

\end{enumerate}

\bibliographystyle{apsrev}
\bibliography{main}

\begin{thebibliography}{75}
\expandafter\ifx\csname natexlab\endcsname\relax\def\natexlab#1{#1}\fi
\expandafter\ifx\csname bibnamefont\endcsname\relax
  \def\bibnamefont#1{#1}\fi
\expandafter\ifx\csname bibfnamefont\endcsname\relax
  \def\bibfnamefont#1{#1}\fi
\expandafter\ifx\csname citenamefont\endcsname\relax
  \def\citenamefont#1{#1}\fi
\expandafter\ifx\csname url\endcsname\relax
  \def\url#1{\texttt{#1}}\fi
\expandafter\ifx\csname urlprefix\endcsname\relax\def\urlprefix{URL }\fi
\providecommand{\bibinfo}[2]{#2}
\providecommand{\eprint}[2][]{\url{#2}}

\bibitem[{\citenamefont{Heeger et~al.}(1988)\citenamefont{Heeger, Kivelson,
  Schrieffer, and Su}}]{SuSchriefferHeeger}
\bibinfo{author}{\bibfnamefont{A.~J.} \bibnamefont{Heeger}},
  \bibinfo{author}{\bibfnamefont{S.}~\bibnamefont{Kivelson}},
  \bibinfo{author}{\bibfnamefont{J.~R.} \bibnamefont{Schrieffer}},
  \bibnamefont{and} \bibinfo{author}{\bibfnamefont{W.~P.} \bibnamefont{Su}},
  \bibinfo{journal}{Rev. Mod. Phys.} \textbf{\bibinfo{volume}{60}},
  \bibinfo{pages}{781} (\bibinfo{year}{1988}).

\bibitem[{\citenamefont{Laughlin}(1983)}]{PhysRevLett.50.1395}
\bibinfo{author}{\bibfnamefont{R.~B.} \bibnamefont{Laughlin}},
  \bibinfo{journal}{Phys. Rev. Lett.} \textbf{\bibinfo{volume}{50}},
  \bibinfo{pages}{1395} (\bibinfo{year}{1983}).

\bibitem[{\citenamefont{Balents}(2010)}]{Balents_Nature}
\bibinfo{author}{\bibfnamefont{L.}~\bibnamefont{Balents}},
  \bibinfo{journal}{Nature} \textbf{\bibinfo{volume}{464}}, \bibinfo{pages}{199
  EP } (\bibinfo{year}{2010}).

\bibitem[{\citenamefont{Zhou et~al.}(2017)\citenamefont{Zhou, Kanoda, and
  Ng}}]{RevModPhys.89.025003}
\bibinfo{author}{\bibfnamefont{Y.}~\bibnamefont{Zhou}},
  \bibinfo{author}{\bibfnamefont{K.}~\bibnamefont{Kanoda}}, \bibnamefont{and}
  \bibinfo{author}{\bibfnamefont{T.-K.} \bibnamefont{Ng}},
  \bibinfo{journal}{Rev. Mod. Phys.} \textbf{\bibinfo{volume}{89}},
  \bibinfo{pages}{025003} (\bibinfo{year}{2017}).

\bibitem[{\citenamefont{Wen}(1991)}]{wen91}
\bibinfo{author}{\bibfnamefont{X.-G.} \bibnamefont{Wen}},
  \bibinfo{journal}{Phys. Rev. B} \textbf{\bibinfo{volume}{44}},
  \bibinfo{pages}{2664} (\bibinfo{year}{1991}).

\bibitem[{\citenamefont{Read and Sachdev}(1991)}]{readsachdev91}
\bibinfo{author}{\bibfnamefont{N.}~\bibnamefont{Read}} \bibnamefont{and}
  \bibinfo{author}{\bibfnamefont{S.}~\bibnamefont{Sachdev}},
  \bibinfo{journal}{Phys. Rev. Lett.} \textbf{\bibinfo{volume}{66}},
  \bibinfo{pages}{1773} (\bibinfo{year}{1991}).

\bibitem[{\citenamefont{Kalmeyer and Laughlin}(1987)}]{kalmeyerlaughlin87}
\bibinfo{author}{\bibfnamefont{V.}~\bibnamefont{Kalmeyer}} \bibnamefont{and}
  \bibinfo{author}{\bibfnamefont{R.~B.} \bibnamefont{Laughlin}},
  \bibinfo{journal}{Phys. Rev. Lett.} \textbf{\bibinfo{volume}{59}},
  \bibinfo{pages}{2095} (\bibinfo{year}{1987}).

\bibitem[{\citenamefont{Kashima and Imada}(2001)}]{kashima01}
\bibinfo{author}{\bibfnamefont{T.}~\bibnamefont{Kashima}} \bibnamefont{and}
  \bibinfo{author}{\bibfnamefont{M.}~\bibnamefont{Imada}}, \bibinfo{journal}{J.
  Phys. Soc. Jpn.} \textbf{\bibinfo{volume}{70}}, \bibinfo{pages}{3052}
  (\bibinfo{year}{2001}).

\bibitem[{\citenamefont{Morita et~al.}(2002)\citenamefont{Morita, Watanabe, and
  Imada}}]{morita02}
\bibinfo{author}{\bibfnamefont{H.}~\bibnamefont{Morita}},
  \bibinfo{author}{\bibfnamefont{S.}~\bibnamefont{Watanabe}}, \bibnamefont{and}
  \bibinfo{author}{\bibfnamefont{M.}~\bibnamefont{Imada}}, \bibinfo{journal}{J.
  Phys. Soc. Jpn.} \textbf{\bibinfo{volume}{71}}, \bibinfo{pages}{2109}
  (\bibinfo{year}{2002}).

\bibitem[{\citenamefont{Lee and Lee}(2005)}]{lee05}
\bibinfo{author}{\bibfnamefont{S.-S.} \bibnamefont{Lee}} \bibnamefont{and}
  \bibinfo{author}{\bibfnamefont{P.~A.} \bibnamefont{Lee}},
  \bibinfo{journal}{Phys. Rev. Lett.} \textbf{\bibinfo{volume}{95}},
  \bibinfo{pages}{036403} (\bibinfo{year}{2005}).

\bibitem[{\citenamefont{Mizusaki and Imada}(2006)}]{mizusaki06}
\bibinfo{author}{\bibfnamefont{T.}~\bibnamefont{Mizusaki}} \bibnamefont{and}
  \bibinfo{author}{\bibfnamefont{M.}~\bibnamefont{Imada}},
  \bibinfo{journal}{Phys. Rev. B} \textbf{\bibinfo{volume}{74}},
  \bibinfo{pages}{014421} (\bibinfo{year}{2006}).

\bibitem[{\citenamefont{Lee et~al.}(2006)\citenamefont{Lee, Nagaosa, and
  Wen}}]{nagaosalee06}
\bibinfo{author}{\bibfnamefont{P.~A.} \bibnamefont{Lee}},
  \bibinfo{author}{\bibfnamefont{N.}~\bibnamefont{Nagaosa}}, \bibnamefont{and}
  \bibinfo{author}{\bibfnamefont{X.-G.} \bibnamefont{Wen}},
  \bibinfo{journal}{Rev. Mod. Phys.} \textbf{\bibinfo{volume}{78}},
  \bibinfo{pages}{17} (\bibinfo{year}{2006}).

\bibitem[{\citenamefont{Balents et~al.}(1998)\citenamefont{Balents, Fisher, and
  Nayak}}]{balents98}
\bibinfo{author}{\bibfnamefont{L.}~\bibnamefont{Balents}},
  \bibinfo{author}{\bibfnamefont{M.~P.~A.} \bibnamefont{Fisher}},
  \bibnamefont{and} \bibinfo{author}{\bibfnamefont{C.}~\bibnamefont{Nayak}},
  \bibinfo{journal}{Int. J. Mod. Phys.} \textbf{\bibinfo{volume}{12}},
  \bibinfo{pages}{1033} (\bibinfo{year}{1998}).

\bibitem[{\citenamefont{Senthil and Fisher}(2000)}]{senthil00}
\bibinfo{author}{\bibfnamefont{T.}~\bibnamefont{Senthil}} \bibnamefont{and}
  \bibinfo{author}{\bibfnamefont{M.~P.~A.} \bibnamefont{Fisher}},
  \bibinfo{journal}{Phys. Rev. B} \textbf{\bibinfo{volume}{62}},
  \bibinfo{pages}{7850} (\bibinfo{year}{2000}).

\bibitem[{\citenamefont{Nomura et~al.}(2017)\citenamefont{Nomura, Darmawan,
  Yamaji, and Imada}}]{PhysRevB.96.205152}
\bibinfo{author}{\bibfnamefont{Y.}~\bibnamefont{Nomura}},
  \bibinfo{author}{\bibfnamefont{A.~S.} \bibnamefont{Darmawan}},
  \bibinfo{author}{\bibfnamefont{Y.}~\bibnamefont{Yamaji}}, \bibnamefont{and}
  \bibinfo{author}{\bibfnamefont{M.}~\bibnamefont{Imada}},
  \bibinfo{journal}{Phys. Rev. B} \textbf{\bibinfo{volume}{96}},
  \bibinfo{pages}{205152} (\bibinfo{year}{2017}).

\bibitem[{\citenamefont{Kaul}(2015)}]{PhysRevLett.115.157202}
\bibinfo{author}{\bibfnamefont{R.~K.} \bibnamefont{Kaul}},
  \bibinfo{journal}{Phys. Rev. Lett.} \textbf{\bibinfo{volume}{115}},
  \bibinfo{pages}{157202} (\bibinfo{year}{2015}).

\bibitem[{\citenamefont{Nomura}(1995)}]{kiyohide.nomura95}
\bibinfo{author}{\bibfnamefont{K.}~\bibnamefont{Nomura}}, \bibinfo{journal}{J.
  Phys. A} \textbf{\bibinfo{volume}{28}}, \bibinfo{pages}{5451}
  (\bibinfo{year}{1995}).

\bibitem[{\citenamefont{Chandra and Doucot}(1988)}]{Chandra_1988}
\bibinfo{author}{\bibfnamefont{P.}~\bibnamefont{Chandra}} \bibnamefont{and}
  \bibinfo{author}{\bibfnamefont{B.}~\bibnamefont{Doucot}},
  \bibinfo{journal}{Phys. Rev. B} \textbf{\bibinfo{volume}{38}},
  \bibinfo{pages}{9335} (\bibinfo{year}{1988}).

\bibitem[{\citenamefont{Capriotti et~al.}(2001)\citenamefont{Capriotti, Becca,
  Parola, and Sorella}}]{Capriotti_2001}
\bibinfo{author}{\bibfnamefont{L.}~\bibnamefont{Capriotti}},
  \bibinfo{author}{\bibfnamefont{F.}~\bibnamefont{Becca}},
  \bibinfo{author}{\bibfnamefont{A.}~\bibnamefont{Parola}}, \bibnamefont{and}
  \bibinfo{author}{\bibfnamefont{S.}~\bibnamefont{Sorella}},
  \bibinfo{journal}{Phys. Rev. Lett.} \textbf{\bibinfo{volume}{87}},
  \bibinfo{pages}{097201} (\bibinfo{year}{2001}).

\bibitem[{\citenamefont{Zhang et~al.}(2003)\citenamefont{Zhang, Hu, and
  Yu}}]{Zhang_2003}
\bibinfo{author}{\bibfnamefont{G.-M.} \bibnamefont{Zhang}},
  \bibinfo{author}{\bibfnamefont{H.}~\bibnamefont{Hu}}, \bibnamefont{and}
  \bibinfo{author}{\bibfnamefont{L.}~\bibnamefont{Yu}}, \bibinfo{journal}{Phys.
  Rev. Lett.} \textbf{\bibinfo{volume}{91}}, \bibinfo{pages}{067201}
  (\bibinfo{year}{2003}).

\bibitem[{\citenamefont{Jiang et~al.}(2012)\citenamefont{Jiang, Yao, and
  Balents}}]{jiang12}
\bibinfo{author}{\bibfnamefont{H.-C.} \bibnamefont{Jiang}},
  \bibinfo{author}{\bibfnamefont{H.}~\bibnamefont{Yao}}, \bibnamefont{and}
  \bibinfo{author}{\bibfnamefont{L.}~\bibnamefont{Balents}},
  \bibinfo{journal}{Phys. Rev. B} \textbf{\bibinfo{volume}{86}},
  \bibinfo{pages}{024424} (\bibinfo{year}{2012}).

\bibitem[{\citenamefont{Wang et~al.}(2013)\citenamefont{Wang, Poilblanc, Gu,
  Wen, and Verstraete}}]{Wang_2013}
\bibinfo{author}{\bibfnamefont{L.}~\bibnamefont{Wang}},
  \bibinfo{author}{\bibfnamefont{D.}~\bibnamefont{Poilblanc}},
  \bibinfo{author}{\bibfnamefont{Z.-C.} \bibnamefont{Gu}},
  \bibinfo{author}{\bibfnamefont{X.-G.} \bibnamefont{Wen}}, \bibnamefont{and}
  \bibinfo{author}{\bibfnamefont{F.}~\bibnamefont{Verstraete}},
  \bibinfo{journal}{Phys. Rev. Lett.} \textbf{\bibinfo{volume}{111}},
  \bibinfo{pages}{037202} (\bibinfo{year}{2013}).

\bibitem[{\citenamefont{Hu et~al.}(2013)\citenamefont{Hu, Becca, Parola, and
  Sorella}}]{hu13}
\bibinfo{author}{\bibfnamefont{W.-J.} \bibnamefont{Hu}},
  \bibinfo{author}{\bibfnamefont{F.}~\bibnamefont{Becca}},
  \bibinfo{author}{\bibfnamefont{A.}~\bibnamefont{Parola}}, \bibnamefont{and}
  \bibinfo{author}{\bibfnamefont{S.}~\bibnamefont{Sorella}},
  \bibinfo{journal}{Phys. Rev. B} \textbf{\bibinfo{volume}{88}},
  \bibinfo{pages}{060402(R)} (\bibinfo{year}{2013}).

\bibitem[{\citenamefont{Qi and Gu}(2014)}]{Qi_2014}
\bibinfo{author}{\bibfnamefont{Y.}~\bibnamefont{Qi}} \bibnamefont{and}
  \bibinfo{author}{\bibfnamefont{Z.-C.} \bibnamefont{Gu}},
  \bibinfo{journal}{Phys. Rev. B} \textbf{\bibinfo{volume}{89}},
  \bibinfo{pages}{235122} (\bibinfo{year}{2014}).

\bibitem[{\citenamefont{Gong et~al.}(2014)\citenamefont{Gong, Zhu, Sheng,
  Motrunich, and Fisher}}]{gong14}
\bibinfo{author}{\bibfnamefont{S.-S.} \bibnamefont{Gong}},
  \bibinfo{author}{\bibfnamefont{W.}~\bibnamefont{Zhu}},
  \bibinfo{author}{\bibfnamefont{D.~N.} \bibnamefont{Sheng}},
  \bibinfo{author}{\bibfnamefont{O.~I.} \bibnamefont{Motrunich}},
  \bibnamefont{and} \bibinfo{author}{\bibfnamefont{M.~P.~A.}
  \bibnamefont{Fisher}}, \bibinfo{journal}{Phys. Rev. Lett.}
  \textbf{\bibinfo{volume}{113}}, \bibinfo{pages}{027201}
  (\bibinfo{year}{2014}).

\bibitem[{\citenamefont{{Richter, Johannes}
  et~al.}(2015)\citenamefont{{Richter, Johannes}, {Zinke, Ronald}, and
  {Farnell, Damian J.J.}}}]{Richter_2015}
\bibinfo{author}{\bibnamefont{{Richter, Johannes}}},
  \bibinfo{author}{\bibnamefont{{Zinke, Ronald}}}, \bibnamefont{and}
  \bibinfo{author}{\bibnamefont{{Farnell, Damian J.J.}}},
  \bibinfo{journal}{Eur. Phys. J. B} \textbf{\bibinfo{volume}{88}},
  \bibinfo{pages}{2} (\bibinfo{year}{2015}).

\bibitem[{\citenamefont{Morita et~al.}(2015)\citenamefont{Morita, Kaneko, and
  Imada}}]{Morita_2015}
\bibinfo{author}{\bibfnamefont{S.}~\bibnamefont{Morita}},
  \bibinfo{author}{\bibfnamefont{R.}~\bibnamefont{Kaneko}}, \bibnamefont{and}
  \bibinfo{author}{\bibfnamefont{M.}~\bibnamefont{Imada}}, \bibinfo{journal}{J.
  Phys. Soc. Jpn.} \textbf{\bibinfo{volume}{84}}, \bibinfo{pages}{024720}
  (\bibinfo{year}{2015}).

\bibitem[{\citenamefont{Wang et~al.}(2016)\citenamefont{Wang, Gu, Verstraete,
  and Wen}}]{Wang_2016}
\bibinfo{author}{\bibfnamefont{L.}~\bibnamefont{Wang}},
  \bibinfo{author}{\bibfnamefont{Z.-C.} \bibnamefont{Gu}},
  \bibinfo{author}{\bibfnamefont{F.}~\bibnamefont{Verstraete}},
  \bibnamefont{and} \bibinfo{author}{\bibfnamefont{X.-G.} \bibnamefont{Wen}},
  \bibinfo{journal}{Phys. Rev. B} \textbf{\bibinfo{volume}{94}},
  \bibinfo{pages}{075143} (\bibinfo{year}{2016}).

\bibitem[{\citenamefont{Poilblanc and Mambrini}(2017)}]{Poilblanc_2017}
\bibinfo{author}{\bibfnamefont{D.}~\bibnamefont{Poilblanc}} \bibnamefont{and}
  \bibinfo{author}{\bibfnamefont{M.}~\bibnamefont{Mambrini}},
  \bibinfo{journal}{Phys. Rev. B} \textbf{\bibinfo{volume}{96}},
  \bibinfo{pages}{014414} (\bibinfo{year}{2017}).

\bibitem[{\citenamefont{Haghshenas and Sheng}(2018)}]{Haghshenas_2018}
\bibinfo{author}{\bibfnamefont{R.}~\bibnamefont{Haghshenas}} \bibnamefont{and}
  \bibinfo{author}{\bibfnamefont{D.~N.} \bibnamefont{Sheng}},
  \bibinfo{journal}{Phys. Rev. B} \textbf{\bibinfo{volume}{97}},
  \bibinfo{pages}{174408} (\bibinfo{year}{2018}).

\bibitem[{\citenamefont{Liu et~al.}(2018)\citenamefont{Liu, Dong, Wang, Han,
  An, Guo, and He}}]{Liu_2018}
\bibinfo{author}{\bibfnamefont{W.-Y.} \bibnamefont{Liu}},
  \bibinfo{author}{\bibfnamefont{S.}~\bibnamefont{Dong}},
  \bibinfo{author}{\bibfnamefont{C.}~\bibnamefont{Wang}},
  \bibinfo{author}{\bibfnamefont{Y.}~\bibnamefont{Han}},
  \bibinfo{author}{\bibfnamefont{H.}~\bibnamefont{An}},
  \bibinfo{author}{\bibfnamefont{G.-C.} \bibnamefont{Guo}}, \bibnamefont{and}
  \bibinfo{author}{\bibfnamefont{L.}~\bibnamefont{He}}, \bibinfo{journal}{Phys.
  Rev. B} \textbf{\bibinfo{volume}{98}}, \bibinfo{pages}{241109}
  (\bibinfo{year}{2018}).

\bibitem[{\citenamefont{Wang and Sandvik}(2018)}]{Wang_2018}
\bibinfo{author}{\bibfnamefont{L.}~\bibnamefont{Wang}} \bibnamefont{and}
  \bibinfo{author}{\bibfnamefont{A.~W.} \bibnamefont{Sandvik}},
  \bibinfo{journal}{Phys. Rev. Lett.} \textbf{\bibinfo{volume}{121}},
  \bibinfo{pages}{107202} (\bibinfo{year}{2018}).

\bibitem[{\citenamefont{Smolensky}(1986)}]{RBM_Smolensky}
\bibinfo{author}{\bibfnamefont{P.}~\bibnamefont{Smolensky}},
  \emph{\bibinfo{title}{Parallel Distributed Processing: Explorations in the
  Microstructure of Cognition: Foundations}} (\bibinfo{publisher}{MIT Press},
  \bibinfo{address}{Cambridge}, \bibinfo{year}{1986}).

\bibitem[{\citenamefont{Carleo and Troyer}(2017)}]{Carleo602}
\bibinfo{author}{\bibfnamefont{G.}~\bibnamefont{Carleo}} \bibnamefont{and}
  \bibinfo{author}{\bibfnamefont{M.}~\bibnamefont{Troyer}},
  \bibinfo{journal}{Science} \textbf{\bibinfo{volume}{355}},
  \bibinfo{pages}{602} (\bibinfo{year}{2017}), ISSN \bibinfo{issn}{0036-8075}.

\bibitem[{\citenamefont{Deng et~al.}(2017{\natexlab{a}})\citenamefont{Deng, Li,
  and Das~Sarma}}]{PhysRevX.7.021021}
\bibinfo{author}{\bibfnamefont{D.-L.} \bibnamefont{Deng}},
  \bibinfo{author}{\bibfnamefont{X.}~\bibnamefont{Li}}, \bibnamefont{and}
  \bibinfo{author}{\bibfnamefont{S.}~\bibnamefont{Das~Sarma}},
  \bibinfo{journal}{Phys. Rev. X} \textbf{\bibinfo{volume}{7}},
  \bibinfo{pages}{021021} (\bibinfo{year}{2017}{\natexlab{a}}).

\bibitem[{\citenamefont{Deng et~al.}(2017{\natexlab{b}})\citenamefont{Deng, Li,
  and Das~Sarma}}]{PhysRevB.96.195145}
\bibinfo{author}{\bibfnamefont{D.-L.} \bibnamefont{Deng}},
  \bibinfo{author}{\bibfnamefont{X.}~\bibnamefont{Li}}, \bibnamefont{and}
  \bibinfo{author}{\bibfnamefont{S.}~\bibnamefont{Das~Sarma}},
  \bibinfo{journal}{Phys. Rev. B} \textbf{\bibinfo{volume}{96}},
  \bibinfo{pages}{195145} (\bibinfo{year}{2017}{\natexlab{b}}).

\bibitem[{\citenamefont{Chen et~al.}(2018)\citenamefont{Chen, Cheng, Xie, Wang,
  and Xiang}}]{PhysRevB.97.085104}
\bibinfo{author}{\bibfnamefont{J.}~\bibnamefont{Chen}},
  \bibinfo{author}{\bibfnamefont{S.}~\bibnamefont{Cheng}},
  \bibinfo{author}{\bibfnamefont{H.}~\bibnamefont{Xie}},
  \bibinfo{author}{\bibfnamefont{L.}~\bibnamefont{Wang}}, \bibnamefont{and}
  \bibinfo{author}{\bibfnamefont{T.}~\bibnamefont{Xiang}},
  \bibinfo{journal}{Phys. Rev. B} \textbf{\bibinfo{volume}{97}},
  \bibinfo{pages}{085104} (\bibinfo{year}{2018}).

\bibitem[{\citenamefont{Glasser et~al.}(2018)\citenamefont{Glasser, Pancotti,
  August, Rodriguez, and Cirac}}]{PhysRevX.8.011006}
\bibinfo{author}{\bibfnamefont{I.}~\bibnamefont{Glasser}},
  \bibinfo{author}{\bibfnamefont{N.}~\bibnamefont{Pancotti}},
  \bibinfo{author}{\bibfnamefont{M.}~\bibnamefont{August}},
  \bibinfo{author}{\bibfnamefont{I.~D.} \bibnamefont{Rodriguez}},
  \bibnamefont{and} \bibinfo{author}{\bibfnamefont{J.~I.} \bibnamefont{Cirac}},
  \bibinfo{journal}{Phys. Rev. X} \textbf{\bibinfo{volume}{8}},
  \bibinfo{pages}{011006} (\bibinfo{year}{2018}).

\bibitem[{\citenamefont{Clark}(2018)}]{1751-8121-51-13-135301}
\bibinfo{author}{\bibfnamefont{S.~R.} \bibnamefont{Clark}},
  \bibinfo{journal}{Journal of Physics A: Mathematical and Theoretical}
  \textbf{\bibinfo{volume}{51}}, \bibinfo{pages}{135301}
  (\bibinfo{year}{2018}).

\bibitem[{\citenamefont{Kaubruegger et~al.}(2018)\citenamefont{Kaubruegger,
  Pastori, and Budich}}]{PhysRevB.97.195136}
\bibinfo{author}{\bibfnamefont{R.}~\bibnamefont{Kaubruegger}},
  \bibinfo{author}{\bibfnamefont{L.}~\bibnamefont{Pastori}}, \bibnamefont{and}
  \bibinfo{author}{\bibfnamefont{J.~C.} \bibnamefont{Budich}},
  \bibinfo{journal}{Phys. Rev. B} \textbf{\bibinfo{volume}{97}},
  \bibinfo{pages}{195136} (\bibinfo{year}{2018}).

\bibitem[{\citenamefont{Lu et~al.}(2019)\citenamefont{Lu, Gao, and
  Duan}}]{PhysRevB.99.155136}
\bibinfo{author}{\bibfnamefont{S.}~\bibnamefont{Lu}},
  \bibinfo{author}{\bibfnamefont{X.}~\bibnamefont{Gao}}, \bibnamefont{and}
  \bibinfo{author}{\bibfnamefont{L.-M.} \bibnamefont{Duan}},
  \bibinfo{journal}{Phys. Rev. B} \textbf{\bibinfo{volume}{99}},
  \bibinfo{pages}{155136} (\bibinfo{year}{2019}).

\bibitem[{Hua()}]{Huang_arXiv}
\bibinfo{note}{Y. Huang, and J. E. Moore, arXiv:1701.06246}.

\bibitem[{\citenamefont{Vieijra et~al.}(2020)\citenamefont{Vieijra, Casert,
  Nys, De~Neve, Haegeman, Ryckebusch, and Verstraete}}]{Vieijra_2020}
\bibinfo{author}{\bibfnamefont{T.}~\bibnamefont{Vieijra}},
  \bibinfo{author}{\bibfnamefont{C.}~\bibnamefont{Casert}},
  \bibinfo{author}{\bibfnamefont{J.}~\bibnamefont{Nys}},
  \bibinfo{author}{\bibfnamefont{W.}~\bibnamefont{De~Neve}},
  \bibinfo{author}{\bibfnamefont{J.}~\bibnamefont{Haegeman}},
  \bibinfo{author}{\bibfnamefont{J.}~\bibnamefont{Ryckebusch}},
  \bibnamefont{and}
  \bibinfo{author}{\bibfnamefont{F.}~\bibnamefont{Verstraete}},
  \bibinfo{journal}{Phys. Rev. Lett.} \textbf{\bibinfo{volume}{124}},
  \bibinfo{pages}{097201} (\bibinfo{year}{2020}).

\bibitem[{\citenamefont{Nomura}(2020)}]{Nomura_2020}
\bibinfo{author}{\bibfnamefont{Y.}~\bibnamefont{Nomura}}, \bibinfo{journal}{J.
  Phys. Soc. Jpn.} \textbf{\bibinfo{volume}{89}}, \bibinfo{pages}{054706}
  (\bibinfo{year}{2020}).

\bibitem[{\citenamefont{Carleo et~al.}(2019)\citenamefont{Carleo, Cirac,
  Cranmer, Daudet, Schuld, Tishby, Vogt-Maranto, and
  Zdeborov\'a}}]{Carleo_2019}
\bibinfo{author}{\bibfnamefont{G.}~\bibnamefont{Carleo}},
  \bibinfo{author}{\bibfnamefont{I.}~\bibnamefont{Cirac}},
  \bibinfo{author}{\bibfnamefont{K.}~\bibnamefont{Cranmer}},
  \bibinfo{author}{\bibfnamefont{L.}~\bibnamefont{Daudet}},
  \bibinfo{author}{\bibfnamefont{M.}~\bibnamefont{Schuld}},
  \bibinfo{author}{\bibfnamefont{N.}~\bibnamefont{Tishby}},
  \bibinfo{author}{\bibfnamefont{L.}~\bibnamefont{Vogt-Maranto}},
  \bibnamefont{and}
  \bibinfo{author}{\bibfnamefont{L.}~\bibnamefont{Zdeborov\'a}},
  \bibinfo{journal}{Rev. Mod. Phys.} \textbf{\bibinfo{volume}{91}},
  \bibinfo{pages}{045002} (\bibinfo{year}{2019}).

\bibitem[{\citenamefont{Choo et~al.}(2018)\citenamefont{Choo, Carleo, Regnault,
  and Neupert}}]{Choo_2018}
\bibinfo{author}{\bibfnamefont{K.}~\bibnamefont{Choo}},
  \bibinfo{author}{\bibfnamefont{G.}~\bibnamefont{Carleo}},
  \bibinfo{author}{\bibfnamefont{N.}~\bibnamefont{Regnault}}, \bibnamefont{and}
  \bibinfo{author}{\bibfnamefont{T.}~\bibnamefont{Neupert}},
  \bibinfo{journal}{Phys. Rev. Lett.} \textbf{\bibinfo{volume}{121}},
  \bibinfo{pages}{167204} (\bibinfo{year}{2018}).

\bibitem[{\citenamefont{Anderson}(1987)}]{ANDERSON1196}
\bibinfo{author}{\bibfnamefont{P.~W.} \bibnamefont{Anderson}},
  \bibinfo{journal}{Science} \textbf{\bibinfo{volume}{235}},
  \bibinfo{pages}{1196} (\bibinfo{year}{1987}), ISSN \bibinfo{issn}{0036-8075}.

\bibitem[{\citenamefont{Liang}(1990)}]{PhysRevB.42.6555}
\bibinfo{author}{\bibfnamefont{S.}~\bibnamefont{Liang}},
  \bibinfo{journal}{Phys. Rev. B} \textbf{\bibinfo{volume}{42}},
  \bibinfo{pages}{6555} (\bibinfo{year}{1990}).

\bibitem[{\citenamefont{Melko et~al.}(2019)\citenamefont{Melko, Carleo,
  Carrasquilla, and Cirac}}]{Melko_2019}
\bibinfo{author}{\bibfnamefont{R.~G.} \bibnamefont{Melko}},
  \bibinfo{author}{\bibfnamefont{G.}~\bibnamefont{Carleo}},
  \bibinfo{author}{\bibfnamefont{J.}~\bibnamefont{Carrasquilla}},
  \bibnamefont{and} \bibinfo{author}{\bibfnamefont{J.~I.} \bibnamefont{Cirac}},
  \bibinfo{journal}{Nat. Phys.} \textbf{\bibinfo{volume}{15}},
  \bibinfo{pages}{887} (\bibinfo{year}{2019}).

\bibitem[{\citenamefont{Mizusaki and Imada}(2004)}]{Mizusaki}
\bibinfo{author}{\bibfnamefont{T.}~\bibnamefont{Mizusaki}} \bibnamefont{and}
  \bibinfo{author}{\bibfnamefont{M.}~\bibnamefont{Imada}},
  \bibinfo{journal}{Phys. Rev. B} \textbf{\bibinfo{volume}{69}},
  \bibinfo{pages}{125110} (\bibinfo{year}{2004}).

\bibitem[{\citenamefont{Pujari et~al.}(2016)\citenamefont{Pujari, Lang, Murthy,
  and Kaul}}]{PhysRevLett.117.086404}
\bibinfo{author}{\bibfnamefont{S.}~\bibnamefont{Pujari}},
  \bibinfo{author}{\bibfnamefont{T.~C.} \bibnamefont{Lang}},
  \bibinfo{author}{\bibfnamefont{G.}~\bibnamefont{Murthy}}, \bibnamefont{and}
  \bibinfo{author}{\bibfnamefont{R.~K.} \bibnamefont{Kaul}},
  \bibinfo{journal}{Phys. Rev. Lett.} \textbf{\bibinfo{volume}{117}},
  \bibinfo{pages}{086404} (\bibinfo{year}{2016}).

\bibitem[{\citenamefont{Tahara and Imada}(2008)}]{doi:10.1143/JPSJ.77.114701}
\bibinfo{author}{\bibfnamefont{D.}~\bibnamefont{Tahara}} \bibnamefont{and}
  \bibinfo{author}{\bibfnamefont{M.}~\bibnamefont{Imada}}, \bibinfo{journal}{J.
  Phys. Soc. Jpn.} \textbf{\bibinfo{volume}{77}}, \bibinfo{pages}{114701}
  (\bibinfo{year}{2008}).

\bibitem[{\citenamefont{Fisher and Barber}(1972)}]{Fisher_Barber}
\bibinfo{author}{\bibfnamefont{M.~E.} \bibnamefont{Fisher}} \bibnamefont{and}
  \bibinfo{author}{\bibfnamefont{M.~N.} \bibnamefont{Barber}},
  \bibinfo{journal}{Phys. Rev. Lett.} \textbf{\bibinfo{volume}{28}},
  \bibinfo{pages}{1516} (\bibinfo{year}{1972}).

\bibitem[{\citenamefont{Suwa et~al.}(2016)\citenamefont{Suwa, Sen, and
  Sandvik}}]{PhysRevB.94.144416}
\bibinfo{author}{\bibfnamefont{H.}~\bibnamefont{Suwa}},
  \bibinfo{author}{\bibfnamefont{A.}~\bibnamefont{Sen}}, \bibnamefont{and}
  \bibinfo{author}{\bibfnamefont{A.~W.} \bibnamefont{Sandvik}},
  \bibinfo{journal}{Phys. Rev. B} \textbf{\bibinfo{volume}{94}},
  \bibinfo{pages}{144416} (\bibinfo{year}{2016}).

\bibitem[{\citenamefont{Sandvik}(2010)}]{PhysRevLett.104.137204}
\bibinfo{author}{\bibfnamefont{A.~W.} \bibnamefont{Sandvik}},
  \bibinfo{journal}{Phys. Rev. Lett.} \textbf{\bibinfo{volume}{104}},
  \bibinfo{pages}{137204} (\bibinfo{year}{2010}).

\bibitem[{\citenamefont{Ferrari and Becca}(2020)}]{Ferrari_2020}
\bibinfo{author}{\bibfnamefont{F.}~\bibnamefont{Ferrari}} \bibnamefont{and}
  \bibinfo{author}{\bibfnamefont{F.}~\bibnamefont{Becca}},
  \bibinfo{journal}{Phys. Rev. B} \textbf{\bibinfo{volume}{102}},
  \bibinfo{pages}{014417} (\bibinfo{year}{2020}).

\bibitem[{\citenamefont{Shao et~al.}(2017)\citenamefont{Shao, Qin, Capponi,
  Chesi, Meng, and Sandvik}}]{Shao_2017}
\bibinfo{author}{\bibfnamefont{H.}~\bibnamefont{Shao}},
  \bibinfo{author}{\bibfnamefont{Y.~Q.} \bibnamefont{Qin}},
  \bibinfo{author}{\bibfnamefont{S.}~\bibnamefont{Capponi}},
  \bibinfo{author}{\bibfnamefont{S.}~\bibnamefont{Chesi}},
  \bibinfo{author}{\bibfnamefont{Z.~Y.} \bibnamefont{Meng}}, \bibnamefont{and}
  \bibinfo{author}{\bibfnamefont{A.~W.} \bibnamefont{Sandvik}},
  \bibinfo{journal}{Phys. Rev. X} \textbf{\bibinfo{volume}{7}},
  \bibinfo{pages}{041072} (\bibinfo{year}{2017}).

\bibitem[{\citenamefont{Harada}(2011)}]{Harada_2011}
\bibinfo{author}{\bibfnamefont{K.}~\bibnamefont{Harada}},
  \bibinfo{journal}{Phys. Rev. E} \textbf{\bibinfo{volume}{84}},
  \bibinfo{pages}{056704} (\bibinfo{year}{2011}).

\bibitem[{\citenamefont{Harada}(2015)}]{Harada_2015}
\bibinfo{author}{\bibfnamefont{K.}~\bibnamefont{Harada}},
  \bibinfo{journal}{Phys. Rev. E} \textbf{\bibinfo{volume}{92}},
  \bibinfo{pages}{012106} (\bibinfo{year}{2015}).

\bibitem[{\citenamefont{Ferrari and Becca}(2018)}]{Ferrari_2018}
\bibinfo{author}{\bibfnamefont{F.}~\bibnamefont{Ferrari}} \bibnamefont{and}
  \bibinfo{author}{\bibfnamefont{F.}~\bibnamefont{Becca}},
  \bibinfo{journal}{Phys. Rev. B} \textbf{\bibinfo{volume}{98}},
  \bibinfo{pages}{100405} (\bibinfo{year}{2018}).

\bibitem[{\citenamefont{Yu et~al.}(2018)\citenamefont{Yu, Wang, Dong, Yao, and
  Li}}]{PhysRevB.98.134410}
\bibinfo{author}{\bibfnamefont{S.-L.} \bibnamefont{Yu}},
  \bibinfo{author}{\bibfnamefont{W.}~\bibnamefont{Wang}},
  \bibinfo{author}{\bibfnamefont{Z.-Y.} \bibnamefont{Dong}},
  \bibinfo{author}{\bibfnamefont{Z.-J.} \bibnamefont{Yao}}, \bibnamefont{and}
  \bibinfo{author}{\bibfnamefont{J.-X.} \bibnamefont{Li}},
  \bibinfo{journal}{Phys. Rev. B} \textbf{\bibinfo{volume}{98}},
  \bibinfo{pages}{134410} (\bibinfo{year}{2018}).

\bibitem[{\citenamefont{Wen}(2002)}]{PhysRevB.65.165113}
\bibinfo{author}{\bibfnamefont{X.-G.} \bibnamefont{Wen}},
  \bibinfo{journal}{Phys. Rev. B} \textbf{\bibinfo{volume}{65}},
  \bibinfo{pages}{165113} (\bibinfo{year}{2002}).

\bibitem[{\citenamefont{Hermele et~al.}(2005)\citenamefont{Hermele, Senthil,
  and Fisher}}]{Hermele_2005}
\bibinfo{author}{\bibfnamefont{M.}~\bibnamefont{Hermele}},
  \bibinfo{author}{\bibfnamefont{T.}~\bibnamefont{Senthil}}, \bibnamefont{and}
  \bibinfo{author}{\bibfnamefont{M.~P.~A.} \bibnamefont{Fisher}},
  \bibinfo{journal}{Phys. Rev. B} \textbf{\bibinfo{volume}{72}},
  \bibinfo{pages}{104404} (\bibinfo{year}{2005}).

\bibitem[{\citenamefont{Xu et~al.}(2019)\citenamefont{Xu, Qi, Zhang, Assaad,
  Xu, and Meng}}]{Xu_2019}
\bibinfo{author}{\bibfnamefont{X.~Y.} \bibnamefont{Xu}},
  \bibinfo{author}{\bibfnamefont{Y.}~\bibnamefont{Qi}},
  \bibinfo{author}{\bibfnamefont{L.}~\bibnamefont{Zhang}},
  \bibinfo{author}{\bibfnamefont{F.~F.} \bibnamefont{Assaad}},
  \bibinfo{author}{\bibfnamefont{C.}~\bibnamefont{Xu}}, \bibnamefont{and}
  \bibinfo{author}{\bibfnamefont{Z.~Y.} \bibnamefont{Meng}},
  \bibinfo{journal}{Phys. Rev. X} \textbf{\bibinfo{volume}{9}},
  \bibinfo{pages}{021022} (\bibinfo{year}{2019}).

\bibitem[{\citenamefont{Misawa et~al.}(2019)\citenamefont{Misawa, Morita,
  Yoshimi, Kawamura, Motoyama, Ido, Ohgoe, Imada, and Kato}}]{MISAWA2019447}
\bibinfo{author}{\bibfnamefont{T.}~\bibnamefont{Misawa}},
  \bibinfo{author}{\bibfnamefont{S.}~\bibnamefont{Morita}},
  \bibinfo{author}{\bibfnamefont{K.}~\bibnamefont{Yoshimi}},
  \bibinfo{author}{\bibfnamefont{M.}~\bibnamefont{Kawamura}},
  \bibinfo{author}{\bibfnamefont{Y.}~\bibnamefont{Motoyama}},
  \bibinfo{author}{\bibfnamefont{K.}~\bibnamefont{Ido}},
  \bibinfo{author}{\bibfnamefont{T.}~\bibnamefont{Ohgoe}},
  \bibinfo{author}{\bibfnamefont{M.}~\bibnamefont{Imada}}, \bibnamefont{and}
  \bibinfo{author}{\bibfnamefont{T.}~\bibnamefont{Kato}},
  \bibinfo{journal}{Comput. Phys. Commun.} \textbf{\bibinfo{volume}{235}},
  \bibinfo{pages}{447 } (\bibinfo{year}{2019}).

\bibitem[{\citenamefont{Sorella}(2001)}]{PhysRevB.64.024512}
\bibinfo{author}{\bibfnamefont{S.}~\bibnamefont{Sorella}},
  \bibinfo{journal}{Phys. Rev. B} \textbf{\bibinfo{volume}{64}},
  \bibinfo{pages}{024512} (\bibinfo{year}{2001}).

\bibitem[{\citenamefont{Choo et~al.}(2019)\citenamefont{Choo, Neupert, and
  Carleo}}]{Choo_2019}
\bibinfo{author}{\bibfnamefont{K.}~\bibnamefont{Choo}},
  \bibinfo{author}{\bibfnamefont{T.}~\bibnamefont{Neupert}}, \bibnamefont{and}
  \bibinfo{author}{\bibfnamefont{G.}~\bibnamefont{Carleo}},
  \bibinfo{journal}{Phys. Rev. B} \textbf{\bibinfo{volume}{100}},
  \bibinfo{pages}{125124} (\bibinfo{year}{2019}).

\bibitem[{\citenamefont{Schulz et~al.}(1996)\citenamefont{Schulz, Ziman, and
  Poilblanc}}]{Schulz_1996}
\bibinfo{author}{\bibfnamefont{H.~J.} \bibnamefont{Schulz}},
  \bibinfo{author}{\bibfnamefont{T.~A.~L.} \bibnamefont{Ziman}},
  \bibnamefont{and}
  \bibinfo{author}{\bibfnamefont{D.}~\bibnamefont{Poilblanc}},
  \bibinfo{journal}{J. Phys. I France} \textbf{\bibinfo{volume}{6}},
  \bibinfo{pages}{675} (\bibinfo{year}{1996}).

\bibitem[{\citenamefont{Liang et~al.}(2018)\citenamefont{Liang, Liu, Lin, Guo,
  Zhang, and He}}]{PhysRevB.98.104426}
\bibinfo{author}{\bibfnamefont{X.}~\bibnamefont{Liang}},
  \bibinfo{author}{\bibfnamefont{W.-Y.} \bibnamefont{Liu}},
  \bibinfo{author}{\bibfnamefont{P.-Z.} \bibnamefont{Lin}},
  \bibinfo{author}{\bibfnamefont{G.-C.} \bibnamefont{Guo}},
  \bibinfo{author}{\bibfnamefont{Y.-S.} \bibnamefont{Zhang}}, \bibnamefont{and}
  \bibinfo{author}{\bibfnamefont{L.}~\bibnamefont{He}}, \bibinfo{journal}{Phys.
  Rev. B} \textbf{\bibinfo{volume}{98}}, \bibinfo{pages}{104426}
  (\bibinfo{year}{2018}).

\bibitem[{\citenamefont{Ferrari et~al.}(2019)\citenamefont{Ferrari, Becca, and
  Carrasquilla}}]{Ferrari_2019}
\bibinfo{author}{\bibfnamefont{F.}~\bibnamefont{Ferrari}},
  \bibinfo{author}{\bibfnamefont{F.}~\bibnamefont{Becca}}, \bibnamefont{and}
  \bibinfo{author}{\bibfnamefont{J.}~\bibnamefont{Carrasquilla}},
  \bibinfo{journal}{Phys. Rev. B} \textbf{\bibinfo{volume}{100}},
  \bibinfo{pages}{125131} (\bibinfo{year}{2019}).

\bibitem[{\citenamefont{Westerhout et~al.}(2020)\citenamefont{Westerhout,
  Astrakhantsev, Tikhonov, Katsnelson, and Bagrov}}]{Westerhout_2020}
\bibinfo{author}{\bibfnamefont{T.}~\bibnamefont{Westerhout}},
  \bibinfo{author}{\bibfnamefont{N.}~\bibnamefont{Astrakhantsev}},
  \bibinfo{author}{\bibfnamefont{K.~S.} \bibnamefont{Tikhonov}},
  \bibinfo{author}{\bibfnamefont{M.~I.} \bibnamefont{Katsnelson}},
  \bibnamefont{and} \bibinfo{author}{\bibfnamefont{A.~A.}
  \bibnamefont{Bagrov}}, \bibinfo{journal}{Nat. Commun.}
  \textbf{\bibinfo{volume}{11}}, \bibinfo{pages}{1593} (\bibinfo{year}{2020}).

\bibitem[{\citenamefont{Szab\'o and Castelnovo}(2020)}]{Szabo_2020}
\bibinfo{author}{\bibfnamefont{A.}~\bibnamefont{Szab\'o}} \bibnamefont{and}
  \bibinfo{author}{\bibfnamefont{C.}~\bibnamefont{Castelnovo}},
  \bibinfo{journal}{Phys. Rev. Research} \textbf{\bibinfo{volume}{2}},
  \bibinfo{pages}{033075} (\bibinfo{year}{2020}).

\bibitem[{Note1()}]{Note1}
Note1, \bibinfo{note}{for large system sizes, we employ sublattice structures
  to make the calculations ``practical'' (making computational cost
  manageable). For example, in Ref. \cite {hu13}, the VMC($p$=2) results are
  not available for large system sizes, and the ``practical'' calculation is
  VMC($p$=1). We notice that, thanks to the retained accuracy, our calculations
  also show state-of-the-art accuracy at the ``practical'' level; for the
  ground-state energy for the $18 \times 18$ lattice at $J_2 = 0.5$, the
  VMC($p$=1) wave function in Ref.~\cite {hu13} gives $E/N_{\protect \rm site}
  = - 0.49611(1)$, whereas our RBM+PP wave function with $6\times 6$ sublattice
  structure gives a better precision of $E/N_{\protect \rm site} =
  -0.496275(3)$}.

\bibitem[{\citenamefont{Nomura}(2021)}]{Nomura_2021}
\bibinfo{author}{\bibfnamefont{Y.}~\bibnamefont{Nomura}}, \bibinfo{journal}{J.
  Phys.: Condens. Matter} \textbf{\bibinfo{volume}{33}},
  \bibinfo{pages}{174003} (\bibinfo{year}{2021}).

\bibitem[{\citenamefont{Kawamura et~al.}(2017)\citenamefont{Kawamura, Yoshimi,
  Misawa, Yamaji, Todo, and Kawashima}}]{KAWAMURA2017180}
\bibinfo{author}{\bibfnamefont{M.}~\bibnamefont{Kawamura}},
  \bibinfo{author}{\bibfnamefont{K.}~\bibnamefont{Yoshimi}},
  \bibinfo{author}{\bibfnamefont{T.}~\bibnamefont{Misawa}},
  \bibinfo{author}{\bibfnamefont{Y.}~\bibnamefont{Yamaji}},
  \bibinfo{author}{\bibfnamefont{S.}~\bibnamefont{Todo}}, \bibnamefont{and}
  \bibinfo{author}{\bibfnamefont{N.}~\bibnamefont{Kawashima}},
  \bibinfo{journal}{Comput. Phys. Commun.} \textbf{\bibinfo{volume}{217}},
  \bibinfo{pages}{180 } (\bibinfo{year}{2017}), ISSN \bibinfo{issn}{0010-4655}.

\end{thebibliography}

\end{document}